\documentclass[a4paper,11pt]{article}

\usepackage{amssymb,amsmath,bm}
\usepackage{a4wide}
\usepackage{color}
\usepackage{slashed}
\usepackage{cite}
\usepackage{graphicx}
\usepackage[latin1]{inputenc}
\usepackage{amsfonts}
\usepackage{lscape}
\usepackage{amsthm}
\usepackage{booktabs}
\usepackage{array}
\usepackage{rotating}
\usepackage{float}
\usepackage{multirow}

\usepackage[small,bf]{caption}
\newcommand{ \mysmall}[1]{\scriptscriptstyle #1} % a smaller #

\newcommand{\GeV}{{\rm GeV}}
\newcommand{\TeV}{{\rm TeV}}
\newcommand{\eps}{\epsilon}
\newcommand{\ord}[1]{\mathcal{O}\left( #1 \right)}

\newcommand{\be}{\begin{equation}}
\newcommand{\ee}{\end{equation}}
\newcommand{\yu}{y_U}
\newcommand{\yd}{y_D}
\newcommand{\ye}{y_E}
\newcommand{\lu}{\lambda_U}
\newcommand{\ld}{\lambda_D}
\newcommand{\lee}{\lambda_E}

\newcommand{\yb}{y_b}
\newcommand{\yt}{y_t}

\newcommand{\yud}{y_U^\dagger}
\newcommand{\ydd}{y_D^\dagger}

\newcommand{\lud}{\lambda_U^\dagger}
\newcommand{\ldd}{\lambda_D^\dagger}
\newcommand{\led}{\lambda_E^\dagger}

\newcommand{\kQQ}{\kappa_{QQ}}

\newcommand{\kUE}{\kappa_{UE}}

\newcommand{\kQL}{\kappa_{QL}}
\newcommand{\kUD}{\kappa_{UD}}

\newcommand{\im}{{\rm Im} }

\selectfont

\begin{document}

\vspace{1cm}
\begin{titlepage}
\vspace*{-1.0truecm}
\begin{flushright}
ULB-TH/13-05 \\
CERN-PH-TH/2013-067 \\
TUM-HEP-882/13 \\
{FLAVOUR(267104)-ERC-39}
\end{flushright}

\vspace{0.8truecm}

\begin{center}
\boldmath

{\Large\textbf{Gauge Mediation beyond Minimal Flavor Violation}}
\unboldmath
\end{center}

\vspace{0.4truecm}

\begin{center}
{\bf Lorenzo Calibbi$^a$, Paride Paradisi$^b$, Robert Ziegler$^{c,d}$}
\vspace{0.4truecm}

{\footnotesize

$^a${\sl Service de Physique Th\'eorique, Universit\'e Libre de Bruxelles, B-1050 Brussels, Belgium}\vspace{0.2truecm}

$^b${\sl CERN, Theory Division, CH-1211 Geneva 23, Switzerland \vspace{0.2truecm}}

$^c${\sl TUM-IAS, Technische Universit\"at M\"unchen, D-85748 Garching, Germany \vspace{0.2truecm}}

$^d${\sl Physik Department, Technische Universit\"at M\"unchen, D-85748 Garching, Germany}

}
\end{center}

%\vspace{0.4cm}
\begin{abstract}
\noindent
We study a minimal modification of Gauge Mediation in which the messenger sector couples directly
to the MSSM matter fields. These couplings are controlled by the same dynamics that explain the
flavor hierarchies, and therefore are parametrically as small as the Yukawas. This setup gives
rise to an interesting SUSY spectrum that is calculable in terms of a single new parameter.
Due to large A-terms, the model can easily accommodate a 126 GeV Higgs with a relatively 
light SUSY spectrum. The flavor structure depends on the particular underlying flavor model, but flavor-violating effects arise dominantly in the up-sector and are strongly suppressed in
$\Delta F =2$ observables.
This strong suppression is reminiscent of what happens in the case of wave function
renormalization or Partial Compositeness, despite the underlying flavor model can be
a simple U(1) flavor model (which in the context of Gravity Mediation suffers from strong
$\Delta F =2$ constraints).
This structure allows to account for the recent observation of direct CP violation in
D-meson decays.

\end{abstract}

\end{titlepage}
%----------------------------------------------------------------------------

%%%%%%%%%%%%%%%%%%%%%%%%%%%%%%%%%%%%%%%%%

\section{Introduction}
While the discovery of low-energy supersymmetry (SUSY) at the LHC is eagerly awaited, the mechanism
of SUSY breaking and its communication to the observable sector still remains unclear. Among the
many candidates, Gauge Mediation  provides an elegant and very predictive framework, since models with gauge-mediated SUSY breaking (GMSB)~\cite{GMSB}  lead to a SUSY spectrum that is completely calculable in terms of few parameters. In the minimal realizations of GMSB sfermion masses are flavor-universal at the messengers scale, so that the only source of flavor violation in the 
sfermion sector are due to renormalization-group (RG) effects from the Standard Model (SM) Yukawa couplings. Therefore these kind of models naturally realize the Minimal Flavor Violation (MFV) paradigm \cite{MFV}, which implies that flavor-violating effects beyond the SM are predicted
to be extremely small.

On the other hand, minimal realizations of GMSB are now seriously challenged \cite{Shih} by
the LHC discovery of a new boson compatible with the SM Higgs, with mass $m_h\approx 126$ GeV~\cite{ATLAS,CMS}.
In the context of the Minimal Supersymmetric Standard Model (MSSM), such a value requires in fact either heavy top squarks (with masses of several TeV) or large left-right stop mixing, in order
to enhance the 1-loop top-stop contribution to $m_h$. In minimal Gauge Mediation A-terms vanish
at leading order at the messenger scale and are only radiatively generated by the RG evolution,
so that the resulting left-right stop mixing is small. Hence, minimal GMSB can account for
$m_h\approx 126$ GeV only at the price of large fine-tuning and a spectrum beyond the reach 
of the LHC. This calls for extensions of the minimal framework, as recently discussed
in Refs.~\cite{Shadmi1, Y1, Y2, Kang, Craig, Babu, Shadmi2, Ray, EvansShih}. For earlier works
in this direction see Refs.~\cite{Dine, Chacko1, Chacko2}.

Most of these scenarios preserve the MFV structure of minimal Gauge Mediation. However, it might
turn out that such a flavor sector is too restrictive and a setup which goes beyond MFV, although
in a controlled way, is favored. An example is provided by the evidence for direct CP violation in $D$ meson decays as reported by the LHCb~\cite{LHCb} and CDF collaborations~\cite{CDF}. Even though at the moment it is not possible to argue that this measurement is a clear signal of physics beyond
the SM~\cite{Golden:1989qx,Brod:2011re,Pirtskhalava:2011va,Cheng:2012wr,
Bhattacharya:2012ah,Li:2012cfa,Franco:2012ck,Feldmann:2012js,Brod:2012ud},
it is interesting to see whether new physics can naturally account for it.

A promising candidate is offered by the framework proposed in Ref.~\cite{Shadmi1}, called ``Flavored Gauge Mediation'' (FGM), where direct couplings of MSSM matter fields to the GMSB messengers are considered. In contrast to e.g. Ref.~\cite{Y1}, where these couplings are directly aligned to the MSSM Yukawa couplings leading to
an MFV scenario, the new couplings share only the same parametric suppression as the Yukawas, but
are not aligned to them.
This flavor structure might result from an underlying theory of flavor that controls Higgs-matter
and messenger-matter in the same way, for example as a consequence of a flavor symmetry under which messengers and Higgs have the same quantum numbers.

In this work we study this framework in great detail and show that it gives rise to an interesting pattern of flavor violation, in which the dominant effects enter through A-terms, i.e. LR mass insertions,
while effects from LL and RR mass insertions are strongly suppressed.
As a result, this framework represents a concrete realization of SUSY models with ``disoriented" A-terms~\cite{GIP}. It is therefore suitable not only to account for the large Higgs mass due to sizable A-terms, but also to address the large amount of direct CP violation in charm consistent
with neutron electric dipole moment (EDM) and $D-\overline{D}$ constraints~\cite{GIP}. We stress that it is highly non-trivial to consider mechanisms of SUSY breaking where flavor violation enters mainly through A-terms. For example flavor models \`a la Froggatt-Nielsen
in the context of gravity mediation can only marginally realize such a setup \cite{Nir}, since they
are challenged by $\Delta F =2$ constraints as a consequence of large off-diagonal entries generated in the LL and RR squark mass matrices. 

The flavor structure of soft terms in FGM in principle depends on the underlying theory of flavor that explains Yukawa hierarchies. However, due to the loop origin of soft terms in GMSB, there is a built-in suppression of $\Delta F=2$ flavor-violating effects so that flavor violation enters dominantly through A-terms, independently of the underlying flavor model. In the special case that the Yukawas and therefore the new messenger--matter couplings have a simple factorizable structure $y\sim\eps_L\eps_R$ as for example in U(1) flavor models, the flavor structure of soft terms resembles that one in SUSY models with Partial Compositeness (PC)~\cite{Nomura, Lodone}. We emphasize that here this particular flavor structure is solely due to the loop origin of soft terms, which acts precisely as a wave function suppression~\cite{Pokorski, Isidori}. In contrast to SUSY PC models this scenario provides complete control of the theory, and therefore allows to study the consequences of perturbative solutions to the 
flavor problem like e.g. flavor symmetries in a very predictive setup for generating soft terms. Indeed, with respect to minimal
gauge mediation, the spectrum of the model is basically controlled by a single new parameter
of the size of the top Yukawa. For a broad range of this parameter the SUSY spectrum is strongly modified with respect to minimal GMSB, with either light stops or light first generation squarks
and gluinos that are potentially observable at the LHC.

Similar studies have already been performed in Refs.~\cite{Y1,Y2} and Ref.~\cite{Shadmi2}. While the conclusions about the SUSY spectrum reproduce the results of Ref.~\cite{Y2}, in this work we try to put a strong emphasis also on the flavor sector which after all is the main new feature of these kinds of models. Based on underlying ideas that were developed in Ref.~\cite{Shadmi1}, in this work we analyze the general flavor structure in great detail and study the corresponding low-energy implications
for flavor physics.

The rest of the paper is organized as follows: After presenting and motivating the model setup
in Section 2 we calculate the new contributions to the soft SUSY breaking terms in Section 3.
In Section 4 we study their consequences for the low-energy SUSY spectrum, while the flavor phenomenology is discussed in Section 5.
In Section 6 we compare the flavor structure of the soft terms of this model to the one of MFV, $U(1)$ flavor models and PC. We conclude in Section 7.

\section{Setup}
We begin with a brief review of Minimal Gauge Mediation. In this scenario $N$ copies of heavy chiral superfields $\Phi_i + \overline{\Phi}_i $ in ${\bf 5+\overline{5}}$ of SU(5) are introduced. These messenger fields couple directly to the SUSY breaking sector, which is effectively parameterized by a single spurion field $X$ that gets a vev $\langle X \rangle = M + F \theta^2$. Through the coupling 
\be
W = X  \overline{\Phi}_i \Phi_i, \qquad i=1 \hdots N
\ee
the messengers acquire large supersymmetric mass terms $M$ and SUSY breaking masses proportional to $F$. By integrating out the messengers at loop-level, soft terms are generated. At the messenger scale, A-terms vanish and gaugino masses and sfermion masses are given by
\begin{gather}
 \label{eq:GMSB-Mi}
 M_i(M)  = N \frac{\alpha_i(M)}{4 \pi} ~\Lambda,  \qquad \qquad \Lambda =  \frac{F}{M}, \\
 \label{eq:GMSB-soft}
 m^2_{\tilde f}(M)  = 2 N \sum_{i=1}^3 C_i(f)~ \frac{\alpha^2_i(M)}{(4 \pi)^2} ~\Lambda^2, \qquad 	f=q,\,u,\,d, \ldots,
\end{gather}
where $C_i(f)$, $i=1,2,3$ is the quadratic Casimir of the representation of the field
$f$ under ${\rm SU(3)}\times {\rm SU(2)}\times {\rm U(1)}$.

Since the messengers have the same gauge quantum numbers as the MSSM Higgs fields,
in addition to the Yukawa couplings
\be
 W = (\yu)_{ij} Q_i U_j H_u +  (\yd)_{ij} Q_i  D_j H_d +  (\ye)_{ij} L_i  E_j H_d,
\ee
also direct couplings of messengers to MSSM fields are allowed by the gauge symmetries.
If we restrict to R-parity even messenger fields, the messengers can couple only to the
MSSM matter fields\footnote{For recent studies of the impact of messenger-matter-Higgs
couplings see Ref.~\cite{Babu, Ray}.}. In general these couplings read
\begin{align}
\label{newcouplgen}
\Delta W & = (\lu)_{ij} Q_i U_j \Phi_{H_u} +  (\ld)_{ij} Q_i  D_j \overline{\Phi}_{H_d} +  (\lee)_{ij} L_i  E_j \overline{\Phi}_{H_d} \nonumber \\
&  + \frac{1}{2} (\kQQ)_{ij} Q_i Q_j \Phi_T + (\kUE)_{ij} U_i E_j \Phi_T \nonumber \\
&  +  (\kQL)_{ij} Q_i L_j \overline{\Phi}_T + (\kUD)_{ij} U_i D_j \overline{\Phi}_T ,
\end{align}
where $\Phi_{H_u}, \Phi_T$ ($\overline{\Phi}_{H_d}, \overline{\Phi}_T$) denote the SU(2) doublet
and SU(3) triplets components of the $\bf 5$ ($\bf \overline{5}$) messenger, and we restricted to
the case of one messenger pair for simplicity.

The presence of direct messenger-matter couplings like in the first line of Eqn.~(\ref{newcouplgen}) gives rise to  new contributions to sfermion masses and A-terms with a flavor structure that depends
on the new parameters $\lambda_{ij}$. If these couplings were flavor-anarchic $\ord{1}$ numbers, the elegant solution of Gauge Mediation to the SUSY flavor problem would be completely spoiled. Therefore it is usually assumed that all direct couplings of the messengers to matter fields vanish, which can
be enforced for example by introducing a new $Z_2$ symmetry under which MSSM fields are even and messengers are odd.

However, there is also another possibility as pointed out in Ref.~\cite{Shadmi1}. Since the
new interactions in the first line of Eqn.~(\ref{newcouplgen}) resemble the MSSM Yukawas,
it is suggestive to assume that also the corresponding couplings are of similar order,
that is, to consider
\be
\label{rel}
\lambda_{U,D,E} \sim y_{U,D,E},
\ee
where $\sim$ denotes equality up to $\ord{1}$ numbers in each entry. Following Ref.~\cite{Shadmi1},
we refer to these kinds of models as ``Flavored Gauge Mediation'' (FGM).
The relation of Eq.~(\ref{rel}) can be justified by assuming that any dynamics that explains the smallness of MSSM Yukawas treats the Higgs fields and the messengers in the same way. For example,
in flavor symmetry models one can assign the same transformation properties to messengers and Higgs fields (in particular that they both transform trivially). Also models with partial compositeness~\cite{Kaplan:1991dc}, where small Yukawas arise from a mixing of the matter superfields with heavy composite states, can serve as a motivation provided that the Higgs fields and the messengers have a similar amount of compositeness (in particular that they are both fully composite).

Since at this point the messengers have the same quantum numbers as the Higgs fields, also their mass terms are given by a  general $2 \times 2$ matrix. This matrix must have a light eigenvalue (the $\mu$-term), and the corresponding light states will be identified with the MSSM Higgs fields, while the heavy eigenstates will be identified with the Gauge Mediation messengers. To explain why one eigenvalue is so light is just the ordinary $\mu-$problem of the MSSM, extended to a $2 \times 2$ matrix. Common solutions to this problem typically introduce new symmetries which forbid the $\mu$-term in the symmetry limit and generate it proportional to the breaking scale. Such symmetries provide a new quantum number that in general allows to differentiate Higgs fields and messengers and will select only a subset of the general couplings in Eq.~(\ref{newcouplgen}). Out of the various possibilities, we focus on the case where only the messenger with the quantum numbers of $H_u$ couples to light fields, that is we consider the superpotential  
\be
\label{newcoupl}
\Delta W = X  \overline{\Phi} \Phi +
(\lu)_{ij} Q_i U_j \Phi_{H_u}
\ee
in addition to the MSSM superpotential. Such a structure can be motivated for example by considering a $U(1)$ symmetry that enforces a zero eigenvalue of the $2 \times 2$ mass matrix of Higgs and messenger fields in the symmetry limit. Since the messengers must be vector-like under this symmetry while the Higgs are chiral, at most one of the messengers can have the same charge as the corresponding Higgs field, and we simply choose to take equal charges for $\Phi_{H_u}$ and $H_u$, as shown in Table 1. We are then free to couple only $\Phi_{H_u}$ to the spurion, which will make it massive with $\overline{\Phi}_{H_d}$. Instead $H_u$ will get a mass term with $H_d$ from the breaking of the $U(1)$. Since this sector of the theory is quite model-dependent, we simply assume that a $\mu$-term of the right size is generated and concentrate on the effects of the new couplings in Eq.~(\ref{newcoupl}). We only take into account that the inclusion of a superpotential term $\Delta W = \mu H_u H_d + \mu^\prime \Phi_{H_u} H_d$ with $\mu \sim \mu^\prime \sim \tilde{m}$ gives a small tree-level correction to $m^2_{H_d}$ that is relevant only for very small messenger scales~\cite{Y1}
\be
\label{treem2Hd}
\Delta m^2_{H_d, tree} = - \frac{\mu^{\prime 2}}{M^2} \frac{\Lambda^2}{1 - \Lambda^2/M^2}.
\ee
Note that we consider a scenario in which the messengers doublets and triplets have different charges. This choice is mainly motivated by following a bottom-up approach, in which we want to restrict to the simplest possibility that gives rise to a large Higgs mass and non-MFV flavor structure. It might be related to the fact that also the MSSM Higgs fields exhibit such an SU(5) breaking structure because in contrast to the Higgs doublets the Higgs triplets must be ultra-heavy. Here we want to require that messenger doublets and triplets have the same mass (coming from the spurion coupling), but they have different couplings to the matter fields as a result of a different $U(1)$ charge assignment. Still, an SU(5) compatible charge assignment does not cause principal problems (note that there is no new source of proton decay since only one Higgs triplet couples to light fields), and will be discussed elsewhere. 
\begin{table}
\centering
\begin{tabular}{c||c|c|c|c|c|c|c|c}
& $\Phi_{H_u}$ &  $\Phi_T$ &  $\overline{\Phi}_{H_d}$ &  $\overline{\Phi}_T$ & $H_u$ & $H_d$ & $X$ & $Q,U,D,E,L$\\
\hline 
$U(1)$ &   $1$ & 0 & $ - 1$ & 0 & 1 & 1 & 0 & $- 1/2$
\end{tabular}
\caption{\label{scens} $U(1)$ charge assignment.}
\end{table}

In summary our setup consists of the superpotential 
\begin{align}
\label{Wfinal}
 W & =   (\yu)_{ij} Q_i U_j H_u +  (\yd)_{ij} Q_i  D_j H_d +  (\ye)_{ij} L_i  E_j H_d \nonumber \\ 
 & + X  \left( \overline{\Phi}_T \Phi_T +  \overline{\Phi}_{H_d} \Phi_{H_u} \right) + (\lu)_{ij} Q_i U_j \Phi_{H_u},  
  \end{align}
together with the assumption that the new parameters $\lambda_U$ are of the same order as the Yukawa couplings
\begin{align}
\label{relationly}
  (\lambda_U)_{ij} \sim (\yu)_{ij}. 
  \end{align}
The superpotential is the most general one consistent with the charge assignment in Table 1 upon redefinition of $\Phi_{H_u}$ and $H_u$. Apart from the new parameters $\lambda_U$ we have the usual parameters of minimal Gauge Mediation, that is $\Lambda \equiv F/M$, the messenger scale $M$ and $\tan{\beta}$. Throughout this paper we will consider only the case of one pair of messengers,
although it is straightforward to generalize this setup to more pairs.

\section{High-energy Spectrum}
We now calculate the SUSY spectrum at the messenger scale. Apart from the usual contributions in Eqs.~(\ref{eq:GMSB-Mi}), (\ref{eq:GMSB-soft}) the presence of the messenger-matter couplings in Eq.~(\ref{Wfinal}) generates new contributions to A-terms and sfermion masses that can be obtained from the general formulae in Ref.~\cite{EvansShih} that are based on the method described in Ref.~\cite{GR}.

In contrast to the minimal setup A-terms arise at 1-loop and are given by
\begin{align}
A_U& = -\frac{\Lambda}{16 \pi^2} \left( \lu \lud \yu +  2 \, \yu \lud \lu \right) \label{eq:Au}\\
A_D & = -\frac{\Lambda}{16 \pi^2} \, \lu \lud \yd \label{eq:Ad} \\
A_E & = 0, \label{eq:Ae}
\end{align}
where all couplings are evaluated at the messenger scale.

Sfermion masses receive new contributions at 1-loop and 2-loop. The 1-loop contributions are suppressed by higher powers of  $x \equiv \Lambda/M$, and thus are relevant only for very low messenger scales. They are given by
\begin{align}
\Delta \tilde{m}^2_{Q, {\rm 1-loop}} & = - \frac{\Lambda^2}{96 \pi^2} x^2 h(x) \lu \lud \label{eq:1loopQ}\\
\Delta \tilde{m}^2_{U, {\rm 1-loop}} & = - \frac{\Lambda^2}{48 \pi^2} x^2 h(x) \lud \lu, \label{eq:1loopU}
\end{align}
with the loop function
\be
h(x) = 3 \, \frac{ (x-2) \log (1-x) - (2+x) \log (1+x) }{x^4} = 1 + \frac{4 x^2}{5} + \ord{x^4}.
\ee
Of course all soft terms involve such loop functions that give sub-leading corrections and can be found in Refs.~\cite{Martin, Y1}. In the parameter space we are considering these corrections to gaugino and 2-loop sfermion masses are negligible, while A-terms receive multiplicative corrections of order $ (1+ \frac{x^2}{3}) $ that we will take into account in the numerical analysis.

The new 2-loop contributions to soft masses read at the messenger scale $M$
\begin{align}
\Delta {\tilde m}^2_E & = \Delta {\tilde m}^2_{L}  =  0, \\
\Delta {\tilde m}^2_{U} & =  \frac{\Lambda^2}{128 \pi^4}
\left[ - \left( \frac{13}{15} g_1^2 + 3 g_2^2 + \frac{16}{3} g_3^2 \right) \lud \lu + \, 3 \, \lud \lu \lud \lu + 3 \, \lud \lu {\rm Tr}  \,  \lu \lud   \right.  \nonumber \\ 
& \left. + \lud \yu \yud \lu + \lud \yd \ydd \lu - \yud \lu \lud \yu + 3 \yud \lu {\rm Tr} \, \yu \lud + 3 \lud \yu {\rm Tr} \, \lu \yud \right] ,\label{eq:m2u}\\
\Delta {\tilde m}^2_{D} & = - \frac{\Lambda^2}{128 \pi^4}  \ydd \lu \lud \yd , \label{eq:m2d}\\
\nonumber \\
\Delta {\tilde m}^2_{Q} & = \frac{\Lambda^2}{256 \pi^4}
\left[  - \left( \frac{13}{15} g_1^2 + 3 g_2^2 + \frac{16}{3} g_3^2 \right) \lu \lud + 3 \, \lu \lud \lu \lud   + 3 \, \lu \lud {\rm Tr} \lu \lud \right. \nonumber \\ & \left. +  \, 2 \, \lu \yud \yu \lud - 2 \, \yu \lud \lu \yud + 3 y_U \lud {\rm Tr} \, \lu \yud + 3 \lu \yud {\rm Tr} \, \yu \lud\right],\label{eq:m2q}\\
\Delta m^2_{H_u} & = -  \frac{3 \Lambda^2}{256 \pi^4} \left[  2 \, {\rm Tr} \, \yu \lud \lu \yud + {\rm Tr} \, \lu \lud \yu \yud \right], \label{eq:m2hu} \\
\Delta m^2_{H_d} & = -   \frac{3 \Lambda^2}{256 \pi^4} {\rm Tr} \, \lu \lud \yd \ydd,  \label{eq:m2hd}
\end{align}
where all couplings are evaluated at the messenger scale. Finally there is the tree-level
contribution to $m^2_{H_d}$ in Eq.~(\ref{treem2Hd}) that arise from the inclusion of the 
$\mu$-term
\be
\Delta m^2_{H_d, tree} = -  \mu^{\prime 2} \frac{x^2}{1 - x^2}.
\ee
In the special case when $(\lu)_{ij} = \lu \delta_{i3} \delta_{j3}$, the above expressions reduce
to the ones obtained in Ref.~\cite{Y1}. Since we will use this approximation in the next section
to calculate the SUSY spectrum, we essentially reproduce the results for the spectrum and the resulting SUSY phenomenology of Ref.~\cite{Y1,Y2}. 

\section{SUSY Phenomenology}
\subsection{General Features of the Low-Energy Spectrum}
We now discuss the consequences for the low-energy spectrum of the new contributions to the
soft terms in Eqs.~(\ref{eq:Au})-(\ref{eq:m2hd}) in the approximation that the only sizable
new coupling is $(\lambda_U)_{33} \equiv \lu$. The spectrum of this model has been previously
studied in \cite{Y1,Y2} and therefore we keep the following discussion rather brief. 

\begin{figure}[t]
\centering
\includegraphics[scale=0.6,angle=-90]{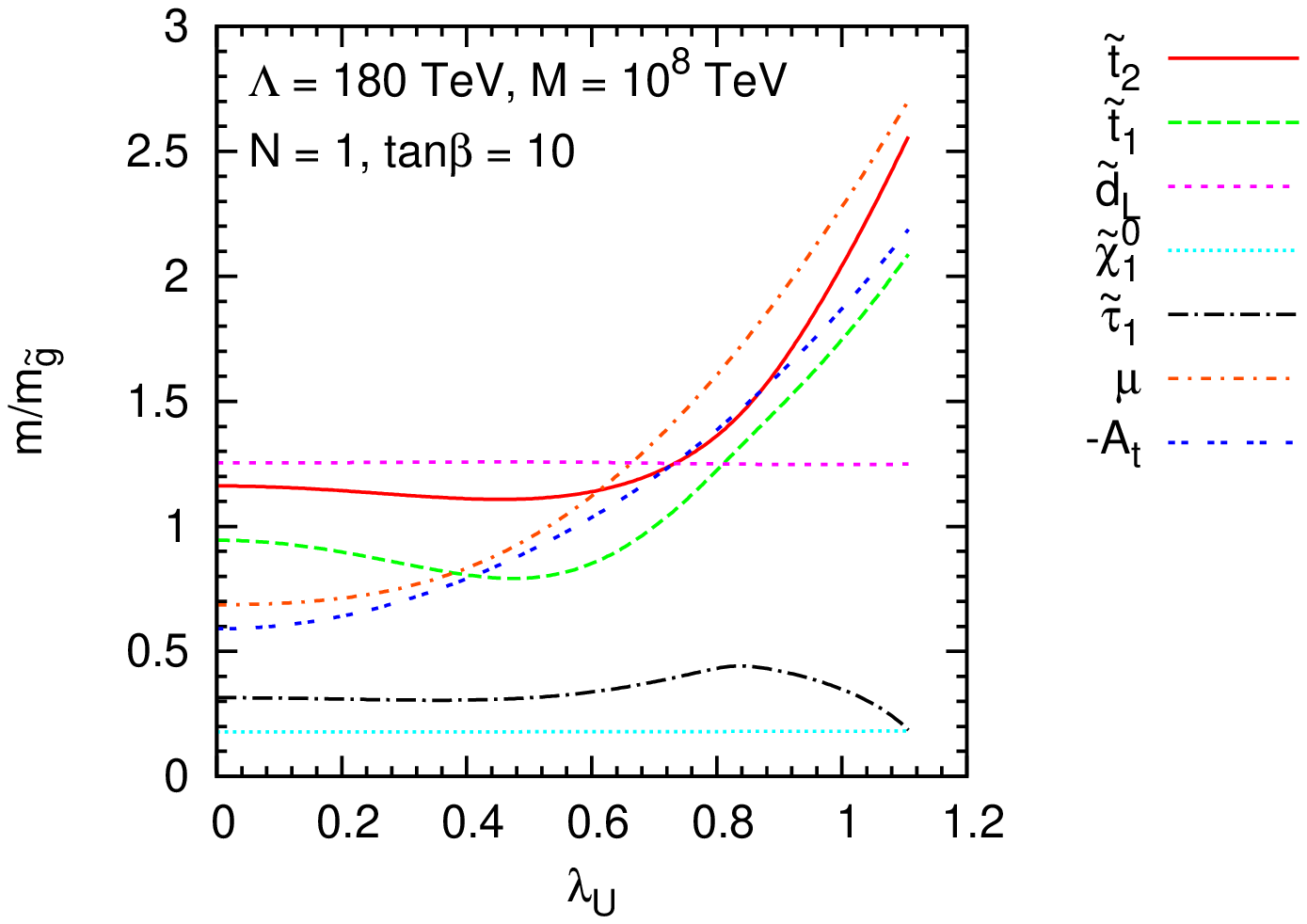}
\includegraphics[scale=0.6,angle=-90]{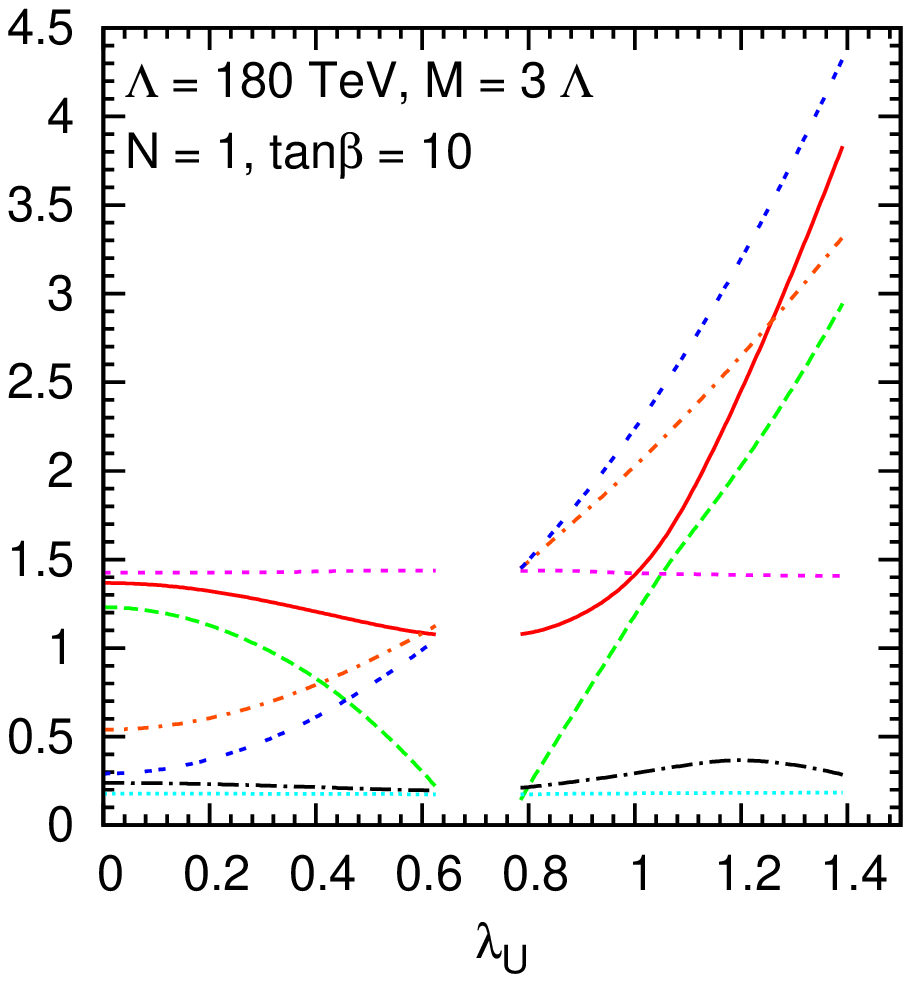}
\caption{\label{specA}Low-energy spectrum normalized by the gluino mass $m_{\tilde g}$. 
The messenger scale is set to $M= 10^8$ TeV (left), $M=3\times\Lambda$ (right).}
\end{figure}

In Fig.~\ref{specA} we plot the low-energy spectrum normalized by
the gluino mass $m_{\tilde g}$ as a function of $\lu$, with $\lambda_U =0 $ corresponding to 
minimal Gauge Mediation.
In the left panel we show the case of a high mediation scale, $M=10^8$ TeV, in the right panel
a low-scale mediation example is displayed, $M = 3\times \Lambda$. The main difference between
the two cases relies on the 1-loop contributions to the stop masses of Eqs.~(\ref{eq:1loopQ},~\ref{eq:1loopU}), that are negligible for $M\gg \Lambda$ but become
relevant in the low-scale mediation case. The other parameters are as indicated in the figure.
The RG running and the SUSY spectrum have been computed by means of the routine {\tt SOFTSUSY} \cite{Allanach:2001kg}.

First of all, we recall that the gaugino spectrum is unchanged with respect to ordinary Gauge Mediation. A-terms instead receive a new negative contribution at the messenger scale, which
grows in absolute value with $\lambda_U$ -- cf.~Eq.~(\ref{eq:Au}) -- and remains sizable at low-energy, as shown in Fig.~\ref{specA}. This has important consequences for the lightest Higgs
boson mass as we will discuss later in more detail. From Eqs.~(\ref{eq:m2hu}) and (\ref{eq:m2hd})
one can see that the new contributions to Higgs soft masses are always negative. This goes in
the same direction as the RG effects, so that Higgs soft masses end up more negative than in
ordinary Gauge Mediation.
This might spoil electroweak symmetry breaking (EWSB), if $m^2_{H_d}$ becomes smaller than $m^2_{H_u}$. 
However, such a situation can occur only in a corner of the parameter space, corresponding to
very large values of $\tan\beta$ (that enhance both the negative high-scale
contribution to $m^2_{H_d}$ in Eq.~(\ref{eq:m2hd}) and the radiative effects), and small values of $A_t$, i.e.~$\lambda_U$ (since otherwise the term in the $m^2_{H_u}$ RGE $\propto |A_t|^2$ always guarantees that $m^2_{H_u}<m^2_{H_d}$).
If this does not happen, then 
$\mu$ and $B \mu$ can be adjusted as usual to allow for successful  EWSB, with $\mu^2 \approx - m^2_{H_u}$ in the largish $\tan{\beta}$ regime. As $m^2_{H_u}$ is more negative, $\mu$ gets
increased and therefore the amount of fine-tuning can be larger than in ordinary Gauge Mediation.
The dependence of $\mu$ on $\lambda_U$ is illustrated in Fig.~\ref{specA}.
For similar reasons, the mass of the CP-odd Higgs, $m_A$, rapidly grows with $\lambda_U$, since $m^2_A \approx  m^2_{H_d}- m^2_{H_u}$.

What regards the stop masses, the new 2-loop contributions can have either sign depending on the relative size of $\lambda_U$ and the gauge couplings, as one can see from Eqs.~(\ref{eq:m2u})
and (\ref{eq:m2q}). For $\lambda_U=0$ the lighter stop is mainly $\tilde{t}_R$, since it does
not receive SU(2) contributions. Switching on $\lambda_U$, for small values the new contribution
to both $\tilde{t}_{L,R}$ is negative, as the negative terms $\propto \lambda_U^2 g_i^2$ dominate
over those $\propto \lambda_U^4$. As a consequence, both stops are lighter than in ordinary Gauge Mediation, as we can see from the left panel of Fig.~\ref{specA}.
Further increasing $\lambda_U$, the stop masses then reach minimal values after which they rise
quickly once $\lambda_U$ is large enough. Since the new contribution to $\tilde{t}_R$ grows
faster with $\lambda_U$ than the one to $\tilde{t}_L$, as can be seen comparing the terms
$\propto \lambda_U^4$ in Eqs.~(\ref{eq:m2u}) and (\ref{eq:m2q}), at some value for $\lambda_U$
the lightest stop becomes mainly left-handed. For $M=\ord{\Lambda}$, the 1-loop contributions of Eqs.~(\ref{eq:1loopQ},~\ref{eq:1loopU}) becomes effective and the stop masses receive a further negative contribution, with the lightest stop possibly getting tachyonic, as in the case shown
in the right panel of the figure.

In contrast to ordinary gauge mediation the sleptons can receive a sizable RG contribution 
\begin{align}
16 \pi^2 \frac{d}{dt} {\tilde m}^2_{L,E} \supset \frac{6}{5} g_1^2  \mathcal{Y}_{L,E} S 
\end{align}
due to the induced hypercharge Fayet--Iliopoulos (FI) term $S$, where $\mathcal{Y}_{L,E}$
is the hypercharge of the $L,~E$ superfields. For large values of $\lambda_U$ the FI-term
is dominated by the stop masses
\begin{align}
S \sim {\tilde m}^2_{t_L} - 2 {\tilde m}^2_{t_R},
\end{align}
and thus is negative in this scenario \cite{Y1,Y2}.
This means that during the growth of $\lambda_U$ the right-handed slepton masses receive
a positive contribution, while the left-handed sleptons are driven lighter.
Since $\tilde{m}^2_E < \tilde{m}^2_L$ for $\lambda_U=0$, the lightest sleptons are mainly
right-handed for small $\lambda_U$, then $\tilde{m}^2_E$ grows and $\tilde{m}^2_L$ decreases
until a turning point, after which lightest sleptons are mainly left-handed.
Since $\tilde{m}^2_L$ continues decreasing with $\lu$, the lightest stau quickly becomes
tachyonic, setting an upper limit for $\lambda_U$.
Such limit is $\lu\simeq 1.2$ in the case depicted in the left panel of Fig.~\ref{specA},
while for a low mediation scale (right panel) the allowed range for $\lu$ is slightly enlarged,
because the ordinary RG contribution to the stau masses $\propto y_\tau^2$ is reduced due
to the smaller length of the running.

The first and second generation squarks do not get new contributions at the high scale, but get large positive RG effects driven by the gluino mass, which are proportional to $\alpha_s$ and therefore particularly strong at small scales. For large $\lu$ the stop masses, and therefore the SUSY breaking scale $M_S\equiv \sqrt{\tilde{m}_{t_1}\tilde{m}_{t_2}}$, are large, which means that the positive gluino effect is less strong than for smaller $\lambda_U$, hence first generation squarks are slightly lighter for larger $\lu$, as shown for the example of ${\tilde d}_L$ in Fig.~\ref{specA}.

Let us finally comment on the dependence of the spectrum on the other parameters: the messenger
index $N$, $\tan{\beta}$ and the messenger scale $M$.\footnote{$\Lambda$ just sets the overall
scale and is therefore not relevant for the normalized spectrum we are showing.}
A larger messenger index $N$ first increases the ordinary contribution to sfermion masses and therefore plays a similar role as $\lambda_U$ in controlling the relative size of ordinary and new contributions. Moreover, since gaugino masses scale with $N$ while sfermion masses with $\sqrt{N}$,
a larger $N$ pushes down the normalized spectrum for small $\lambda_U$. As mentioned above, the messenger scale $M$ influences the length of the RG running and controls the size of the 1-loop contributions to the stop masses. Besides that, it has also some impact on the new 2-loop contributions, since it determines the strength of the gauge couplings. For example $\alpha_s$
grows with smaller $M$, which means that e.g.~the minimum in the stop masses occurs for larger $\lambda_U$ (as can be seen by comparing the two panels of Fig.~\ref{specA}).
The only effect of larger values of $\tan{\beta}$ (besides enhancing the usual negative RG contributions to third generation sfermions, $\propto y_b^2$ and $y_\tau^2$, as well as the
LR sbottom and stau mixing terms) is a slight increase of the $\tilde{t}_R$ mass due to the
positive term $\propto y_b^2 \lambda_U^2$ in Eq.~(\ref{eq:m2u}).

\subsection{Highlights of the Low-energy Spectrum}
\begin{figure}[t]
\centering
\includegraphics[height=0.6\textwidth,angle=-90]{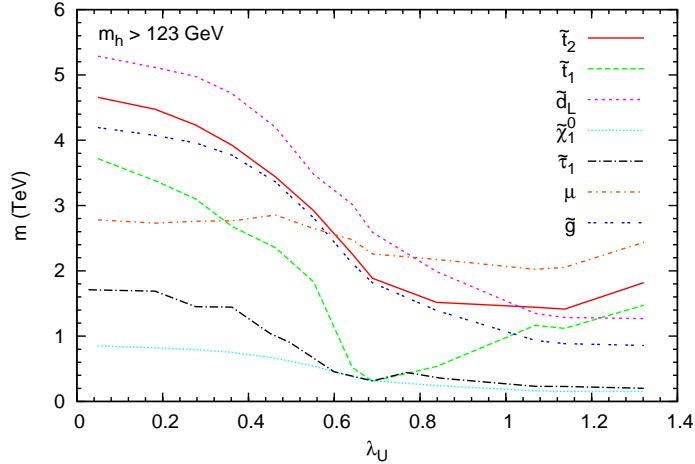}
\caption{\label{spec122} Lower bounds on sparticle masses corresponding
to a Higgs boson mass larger than 123 GeV. Ordinary Gauge Mediation
corresponds to $\lu =0$.}
\end{figure}

We now discuss the main new features of the spectrum described in the last section. The most important difference with respect to ordinary Gauge Mediation is the occurrence of large negative A-terms which can lead to large Higgs boson masses even for a quite light spectrum. First of all, A-terms cannot be too large because they have to respect the vacuum stability bound~\cite{VacStab}.
\begin{align}
A_t^2 + 3 \mu^2 \le 7.5 \, ({\tilde m}^2_{t_L} + {\tilde m}^2_{t_R}).
\label{eq:stab}
\end{align}
If this bound is fulfilled, the A-terms control the top-stop 1-loop contribution to the Higgs mass through the stop mixing parameter $X_t \equiv {A_t}-\mu\, \cot \beta $, which for $\tan{\beta} \gtrsim 10$ is basically given by $A_t$. The approximate expression of this radiative correction reads:
\begin{align}
\Delta m^2_h = \frac{3 m_t^4}{8 \pi^2 v^2} \left( \log\frac{M_S^2}{m_t^2} + \frac{X_t^2}{M_S^2} \left(1 - \frac{X_t^2}{12 M_S^2}\right)\right),
\end{align}
where $M_S\equiv \sqrt{\tilde{m}_{t_1}\tilde{m}_{t_2}}$ and $v \approx \, 174~\GeV$.
This contribution is maximized for $|X_t /M_S| \approx \sqrt{6}$, in which case it brings up the
Higgs mass to $125~\GeV$ for $M_S \sim 1~\TeV$ (see e.g.~\cite{Arbey:2011ab}). As can be seen from Fig.~\ref{specA}, the ratio $|A_t/M_S|$ is always larger than in minimal Gauge Mediation, with the typical value of $|A_t/M_S|\approx 1$ for a high messenger scale (left panel).
In the case of low messenger scale (right panel), $|A_t/M_S|$ can easily reach $\sqrt{6}$, which maximizes the 1-loop contributions to $m_h$. This implies that the average stop mass $M_S$ can be
much lighter compared to minimal Gauge Mediation for the same value of $m_h$.

Requiring a certain Higgs mass then fixes the overall scale of the SUSY spectrum of Fig.~\ref{specA}.
In Fig.~\ref{spec122} we plot the lower bounds of some sparticle masses requiring $m_h > 123$ GeV,
a value compatible with the observed $m_h \approx 126$ GeV, once experimental and theoretical uncertainties are taken into account (the latter can be estimated to be about 3 GeV, see e.g.~\cite{Arbey:2012dq}).
The figure has been obtained performing a scan of the parameters in the following ranges:
\begin{align}
5 \le \tan\beta\le 40, \quad 40~{\rm TeV} \le \Lambda \le 700~{\rm TeV}, \quad
\Lambda < M \le 10^{15}~{\rm GeV}, \quad  N=1, \nonumber 
\end{align}
and selecting the points corresponding to the lowest possible stop masses compatible with
$m_h > 123$ GeV. The Higgs mass has been computed by {\tt SOFTSUSY} \cite{Allanach:2001kg},
as the rest of the spectrum.

Fig.~\ref{spec122} shows that the $\mu$ term, and thus the fine tuning, can be slightly smaller
than in ordinary Gauge Mediation for $m_h>123$ GeV. Moreover, the SUSY particles can be much
lighter. For instance, near the minimum at $\lu \approx 0.7$, the lightest stop can be as light
as 400 GeV. Also larger values of $\lu$ are interesting since they can give relatively light
first generations squarks, gluinos (relevant for LHC searches) and LH sleptons. This latter
feature can be particularly relevant for the SUSY contribution to the anomalous magnetic moment
of the muon, $\Delta a_\mu \equiv (g-2)^{\rm SUSY}_\mu/2$, as pointed out in \cite{Y2}.
\begin{figure}[t]
\centering
\includegraphics[height=0.6\textwidth,angle=-90]{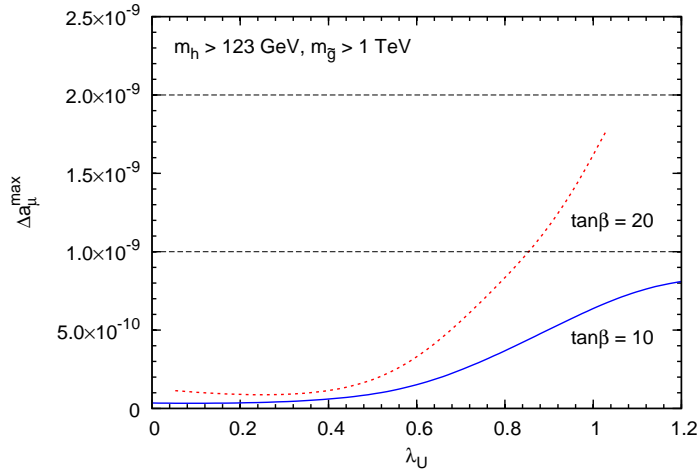}
\caption{\label{gm2}  Largest SUSY contribution to $\Delta a_\mu$
compatible with $m_h>123$ GeV, as a function of $\lambda_U$.}
\end{figure}

As we discussed, LH slepton masses strongly decrease for sizeable values of $\lu$
(and can become smaller than the RH ones) and thus are significantly lighter than
in ordinary Gauge Mediation (where $\tilde{m}_{\mu_L} \gtrsim 2.5~\TeV $ for $m_h> 123$ GeV).
Moreover, for sizeable values of $\lu$, $\mu\gg \tilde{m}_{\mu_L},~\tilde{m}_{\mu_R}$.
In this regime, $\Delta a_\mu$ is dominated by the pure Bino contribution, which is
$\mu$-enhanced, in contrast to the usually dominant $\mu$-suppressed Wino-Higgsino
contribution, as recently discussed in \cite{GIS}.
Taking for simplicity $\tilde{m}_{\mu_L} = \tilde{m}_{\mu_R} = M_1 \equiv {\tilde m}$,
$\Delta a_\mu$ can be approximated by the following expression \cite{GIS}:
\begin{align}
\Delta a_\mu \approx 1 \times 10^{-9} \left(\frac{\tan{\beta}}{20} \right) \left(\frac{500 \, \GeV}{{\tilde m}} \right)^2 \left(\frac{1}{8}\frac{10}{\mu/{\tilde m}} + \frac{\mu/{\tilde m}}{10} \right).
\end{align}
From this, we see that the SUSY contribution to $(g-2)_\mu$ can be easily large enough to lower
the tension with the experimental measurements. As an illustration, we plot in Fig.~\ref{gm2}
$\Delta a_\mu$ (computed using the full expression of~\cite{Moroi:1995yh}) as a function of
$\lu$ for $\tan{\beta} = 10,\,20$, again fixing the lightest spectrum compatible with $m_h > 123\,\GeV$.
As we can see, $\Delta a_\mu$ can reach values of about $(1\div 2)\times 10^{-9}$ if $\lu\approx 1$,
thus reducing significantly the $\sim 3.5 \sigma$ discrepancy between the SM prediction and the experimental value
$\Delta a_\mu =a_\mu^{\mysmall \rm EXP}-a_\mu^{\mysmall \rm SM} = 2.90 (90) \times 10^{-9}$
~\cite{bnl, Jegerlehner:2009ry, HLMNT11, DHMZ11}.
Much larger values of $\tan{\beta}$ are not viable for $\lu \approx 1$ because of
tachyonic staus and thus do not lead to a further enhancement of $\Delta a_\mu$.

We now comment about SUSY searches at the LHC.
Searches for GMSB scenarios depend on the gravitino mass as well as on
the nature of the next-to-lightest SUSY particle (NLSP). Constraints
and prospects for the present model have been discussed in detail in
\cite{Y2}.
For $N=1$, the NLSP is typically a Bino-like neutralino, 
although stau NLSP is
possible especially for large $\lu$, as
follows from the above discussion. Depending on the gravitino mass,
the NLSP can decay promptly or be long-lived. The former case is
realized for very light gravitinos corresponding to $M\lesssim 1000$
TeV.
However, larger F-terms in other sectors of the theory can raise the
gravitino mass even for such low mediation scales. As a consequence, in
our setup the parameters can always be adjusted such that the NLSP is
a long-lived neutralino, decaying outside the detector. In this case
LHC searches resemble the ones for
typical gravity mediation scenarios based on multi-jets and
$\slashed{E}_T$ events.
The most recent ATLAS \cite{SUSYatlas} and CMS \cite{SUSYcms} analysis
then set a bound of about $m_{\tilde g}\gtrsim 1.2$ TeV, corresponding
in our model to first generations squarks with $\tilde{m}_q\gtrsim
1.6$ TeV
and $\tilde{m}_\tau,\, m_{\tilde{\chi}^0}\gtrsim 230$ GeV. Limits in  the case
of a prompt decay $\tilde{\chi}^0\to \tilde{G} \gamma$ \cite{Aad:2012zza} or
long-lived stau \cite{Aad:2012pra} are slightly more stringent (roughly $m_{\tilde
g}\gtrsim 1.5$ TeV). In the case of a promptly decaying stau NLSP, recent searches
based on events with jets, $\slashed{E}_T$ and at least one $\tau$ \cite{Prompt-stau}
set a bound of $m_{\tilde g}\gtrsim 1.2~(1.5)$ TeV for low (large) values of $\tan\beta$.

Finally we notice that $h\to\gamma\gamma$ remains SM-like in this model.
In principle, a light stau (with large left-right mixing) and heavy
higgsinos, could enhance
$h\to\gamma\gamma$ up to $\sim 50\%$ for $m_{{\tilde\tau}_1}\lesssim
100~$GeV~\cite{hggSUSY, GIS}.
As we have seen, in this scenario $m_{{\tilde\tau}_1}\gtrsim 230~$GeV and
therefore we find at most only a few per-cent enhancement.

\section{Flavor Phenomenology}
We now turn to the flavor phenomenology of the model that is determined by the flavor structure
of the $\lambda_U$ couplings. These matrices are assumed to have the same hierarchical structure
as the Yukawa couplings, as a consequence of some underlying theory of flavor.

Such a theory could be based on a flavor symmetry or other rationales like partial compositeness,
and is in principle needed in order to make quantitative predictions in the flavor sector. However
the new flavor-violating effects can be sizable only in a limited sector of the theory, which drastically reduces this ambiguity.
As we will see, sizable effects arise only through $\delta^u_{LR}$ and  $\delta^u_{RR}$, the latter
being strongly suppressed. Such a structure is precisely what one needs in order to account for direct
CP violation in charm decays in the context of SUSY, and indeed one can easily generate a sizable CP
asymmetry given a suitable Yukawa structure. Before presenting the numerical analysis we discuss
the general structure of flavor-violating effects.
\subsection{General Flavor Structure}
From the expressions for soft terms in Eqs.~(\ref{eq:Au})-(\ref{eq:m2hd}) one can see that the
flavor structure of the new contributions takes the form
\begin{align}
A_U & \sim \lu \lud \yu + \yu \lud \lu, & A_D & \sim \lu \lud \yd, 
\end{align}
\begin{align}
\Delta \tilde{m}^2_Q & \sim \lu \lud, & \Delta \tilde{m}^2_U & \sim \lud \lu, & \Delta \tilde{m}^2_D
\sim \ydd \lu \lud \yd. &
\end{align}
Because of the particular underlying loop structure the A-terms and the RH down masses are partially aligned to Yukawa matrices, so they will be suppressed by light Yukawas in the mass basis. If we go
to this basis where Yukawas are diagonalized by bi-unitary transformations
\begin{align}
(V^U_L)^\dagger \yu V^U_R = \yu^{diag} \qquad  \qquad (V^D_L)^\dagger \yd V^D_R = \yd^{diag},
\end{align}
the new couplings transform as
\begin{align}
(V^U_L)^\dagger \lu V^U_R = \hat{\lambda}_U.
\end{align}
The matrices $\hat{\lambda}_U$ can always be calculated given the structure of Yukawa matrices.
In the special case where Yukawas (and therefore $\lambda_U$) are hierarchical, that is
$y_{ij} \le y_{i^\prime j}$ for $i^\prime > i$ and $y_{ij} \le y_{i j^\prime }$ for $j^\prime > j$,
the bi-unitary transformations do not change the hierarchical structure of $\lambda_U$ but only
the $\ord{1}$ coefficients. In the hierarchical case the relation in Eq.~(\ref{relationly})
therefore remains valid also in the mass basis, that is
\be
\hat{\lambda}_U \sim \lu \sim \yu.
\ee
In the fermion mass basis we then obtain for the parametric dependence of flavor violating
mass insertions
\begin{align}
\label{deltaLL1}
(\delta^u_{LL})_{ij}  & \sim  (\lu)_{i3} (\lu^*)_{j3}, &
(\delta^d_{LL})_{ij} & \sim V_{3i} V_{3j}^* y^2_t , \\
\label{deltaRR1}
(\delta^u_{RR})_{ij} & \sim (\lu^*)_{3i} (\lu)_{3j}, & 
(\delta^d_{RR})_{ij} & \sim y^D_i  y^D_j  V_{3i} V_{3j}^* y_t^2,
\end{align}
\begin{gather}
\label{deltaLRup}
(\delta^u_{LR})_{ij} \sim
\frac{A m^U_j}{{\tilde m}_Q{\tilde m}_U} \frac{(\lu)_{i3} (\lu^*)_{j3}}{y^2_t } +
\frac{A m^U_i}{{\tilde m}_Q{\tilde m}_U} \frac{(\lu^*)_{3i} (\lu)_{3j}}{y^2_t }, \\
\label{deltaLRdown}
(\delta^d_{LR})_{ij} \sim \frac{A m^D_j}{{\tilde m}_Q{\tilde m}_D} V_{3i} V_{3j}^*,
\end{gather}
where $V = (V^U_L)^T (V^D_L)^*$ denotes the CKM matrix and $A$ is defined by $(A_U)_{33} = A \, y_t$. We also assumed hierarchical Yukawas (so that $\hat{\lambda}_U \sim \lambda_U$) and have dropped the flavor-diagonal SUSY contribution to the LR sfermion mass matrix for simplicity. Notice that there are no new effects in the slepton sector.

Since the Yukawa entries $(\yu)_{i<j}$ determine the left-handed rotations, in the absence
of cancellations between up- and down sector they are constrained by the CKM matrix, and so
are the $\lambda_U$ couplings,
$(\lu)_{13} \lesssim \lambda^3$, $(\lu)_{23} \lesssim \lambda^2$ ($\lambda \approx 0.2$).
Therefore both $\delta_{LL}^{u}$ and $\delta_{LL}^{d}$ have a CKM suppression, while $\delta_{RR}^d$ and $\delta_{LR}^d$ are negligibly small due to CKM suppression in addition to light Yukawa suppression (this is a consequence of the fact that there is no new spurion transforming under $SU(3)_d$).
Therefore flavor violating effects manifest themselves dominantly through $\delta^u_{RR}$ and $\delta^u_{LR}$. Since the latter is partially aligned to the Yukawas, the prevailing effect
for light generations is through an ``effective" triple LR mass insertion 
$(\delta^u_{LR})_{ij}^{eff} \equiv
(\delta^u_{LL})_{i3}(\delta^u_{LR})_{33} (\delta^u_{RR})_{3j}$ or double  LR mass insertion $(\delta^u_{LR})_{ij}^{eff} \equiv (\delta^u_{LL})_{i3}(\delta^u_{LR})_{3j}$ and $(\delta^u_{LR})_{ij}^{eff} \equiv (\delta^u_{LR})_{i3}(\delta^u_{RR})_{3j}$, since in this
way the diagonal Yukawa coupling can be sandwiched between the $\lambda_U$ spurions that are
not diagonal in the mass basis.  The dominant effect is then given by
\begin{align}
\label{delta2}
(\delta^u_{LR})_{ij}^{eff} \sim
\frac{m_t(A - \yt^2 \mu^*/\tan\beta)}{{\tilde m}_Q{\tilde m}_U}
 \, (\lu)_{i3}(\lu)_{3j}~\qquad i,j = 1,2.
\end{align}
Although the flavor structure of $\lu$ depends on the particular underlying flavor model,
we stress that flavor-violating effects arise dominantly in the up-sector and are strongly
suppressed in $\Delta F =2$ observables. To see this point more explicitly, we restrict to
the case where $\lu$ is controlled by a single $U(1)$ symmetry so that
$(\yu)_{ij} \sim (\lu)_{ij} \sim \eps^{Q_i + U_j}$ where $\eps$ is a small order parameter 
and $Q_i,U_i$ denote the positive $U(1)$ charges. In such a case, we find
\begin{align}
%\label{deltaLL}
(\delta^u_{LL})_{ij}  & \sim V_{i3}^* V_{j3} \yt^2, &
(\delta^u_{RR})_{ij} & \sim \frac{y^U_i y^U_j}{V_{i3}^* V_{j3}},
\end{align}
\begin{equation}
(\delta^u_{LR})_{ij}^{eff} \sim
\frac{m^U_j \left(A - \yt^2 \mu^*/\tan\beta\right)}{{\tilde m}_Q{\tilde m}_U}
\frac{V_{i3}^*}{V_{j3}} \yt^2~\qquad i,j = 1,2.
\end{equation}
Despite the underlying $U(1)$ flavor symmetry, the GMSB setup leads to a strong suppression
for LL and RR mass insertions, which is reminiscent of what happens in the case of wave
function renormalization~\cite{Pokorski, Isidori} or Partial Compositeness~\cite{Nomura,Lodone}.
In FGM models, the extra suppression originates in the loop origin of soft terms, which acts
precisely as a wave function suppression.

This peculiar flavor structure leads to phenomenological consequences that
are are especially important for charm physics and hadronic EDMs, as we are
going to discuss now.
\subsection{Flavor Predictions}
The existence of direct charm CP violation in $D\to P^+P^-$ decays ($P=\pi,K$) has been
firmly established experimentally after the measurements by the LHCb and CDF collaborations.
The combination of the LHCb~\cite{LHCb} and CDF~\cite{CDF}
results with previous measurements leads to the world average~\cite{CDF}
\footnote{Recently, the LHCb collaboration claimed the new result
$\Delta a_{CP} = (0.49 \pm 0.30 \pm 0.14)\%$~\cite{Aaij:2013bra} that has been
obtained using $D^0$ mesons produced in semileptonic b-hadron decays. By contrast,
all previous analyses have used $D^0$ mesons from $D^{*+}\to D^{0}\pi^{+}$ decays.
This result does not confirm the evidence for direct CP violation in the charm sector
reported in all previous analyses. Hereafter, we do not consider the new LHCb result
waiting for its confirmation.}
\begin{equation}
\Delta a_{CP} \equiv a_{K^+ K^-} - a_{\pi^+ \pi^-}  = -(0.68 \pm 0.15)\%\, ,
\label{eq:acpExp}
\end{equation}
where
\begin{equation}
a_f \equiv \frac{\Gamma(D^0\to f)-\Gamma(\bar D^0\to f)}
  {\Gamma(D^0\to f)+\Gamma(\bar D^0\to f)}\, , ~~f=K^+ K^-,\pi^+ \pi^-.
\label{eq:af_def}
\end{equation}
The result in eq.~(\ref{eq:acpExp}) represents an evidence for CP violation
at the $4\sigma$ level.

In the SM it turns out that $\Delta a^{\rm SM}_{CP}\approx -(0.13\%) \times\mathrm{Im}
(\Delta R^{\rm SM})$~\cite{Isidori:2011qw}, where $\Delta R^{\rm SM}$ stands for ratios
of hadronic amplitudes which in perturbation theory are expected to be of order
$\Delta R^{\rm SM} \approx \alpha_s (m_c)/\pi\approx 0.1$, but a significant enhancement
could arise from non-perturbative effects
(see refs.~\cite{Golden:1989qx,Brod:2011re,Pirtskhalava:2011va}).

Therefore, we assume hereafter that new physics effects are at work and we investigate on the implications of this measurement. As throughly discussed in ref.~\cite{Grossman:2006jg,GIP},
new physics theories generating a large CP violating phase for the $\Delta C=1$ chromomagnetic
operator are the best candidate to explain the result in question while naturally accounting
for all current flavor data, especially $D^0-\bar D^0$ mixing data. New-physics (NP) effects
are encoded in the effective Hamiltonian
\be
\mathcal H^{\rm eff-\rm NP}_{|\Delta c| = 1} = \frac{G_F}{\sqrt 2}
  \sum_{i} \left(C_i Q_i + C^{\prime}_{i} Q^{\prime}_{i} + {\rm h.c.}\right)\,,
\label{eq:HNP}
\ee
where the relevant electromagnetic and chromomagnetic dipole operators read
\begin{eqnarray}
Q_{7} &=&  \frac{m_c}{4\pi^2}\, \bar u_L \sigma_{\mu\nu}
           e F^{\mu\nu} c_R \,, \nonumber  \\
Q_{8} &=&  \frac{m_c}{4\pi^2}\, \bar u_L \sigma_{\mu\nu}
           g_s G^{\mu\nu} c_R \,.
   \label{eq:Q8def}
\end{eqnarray}
As usual, $Q^{\prime}_{7,8}$ are obtained from $Q_{7,8}$ by exchanging
$L\leftrightarrow R$.

At the low (physical) scale $m_c$, the expression for $C^{(\prime)}_{7,8}$ can
be obtained from the corresponding expression at high scale taking into account
the leading log RG evolution of the operators~\cite{Buchalla:1995vs}
\begin{eqnarray}
C^{(\prime)}_7(m_c) &=& \tilde \eta  \left[ \eta C^{(\prime)}_7({\tilde m})
+ 8 Q_u\, (\eta -1)\, C^{(\prime)}_8({\tilde m}) \right],
\label{eq:C7_mc}
\\
C^{(\prime)}_8(m_c) &=& \tilde \eta\, C^{(\prime)}_8({\tilde m}),
\label{eq:C8_mc}
\end{eqnarray}
where $Q_u=2/3$ is the up-quark electric charge and
\begin{equation}
\label{eta}
\eta=\left[\frac{\alpha_s({\tilde m})}{\alpha_s(m_t)}\right]^\frac{2}{21}
        \left[\frac{\alpha_s(m_t)}{\alpha_s(m_b)}\right]^\frac{2}{23}
        \left[\frac{\alpha_s(m_b)}{\alpha_s(m_c)}\right]^\frac{2}{25}~,
\end{equation}
\begin{equation}
\tilde \eta = \left[ \frac{\alpha_s({\tilde m})}{\alpha_s(m_t)}\right]^{\frac{14}{21}}
\left[ \frac{\alpha_s(m_t)}{\alpha_s(m_b)}\right]^{\frac{14}{23}}
\left[ \frac{\alpha_s(m_b)}{\alpha_s(m_c)}\right]^{\frac{14}{25}}\,.
\end{equation}
Following the QCD factorization approach of ref.~\cite{Grossman:2006jg}, which we assume
for definiteness keeping in mind that it suffers from ${\mathcal O}(1)$ uncertainties,
one can find that
\begin{equation}
|\Delta a_{CP}| \approx \frac{4}{\sin\theta_c} \frac{\alpha_s (m_c)}{\pi}
\left|\mathrm{Im}\left(C_{8}(m_c) + C^{\prime}_{8}(m_c)\right) \right|\,,
\label{eq:acp_np}
\end{equation}
where, hereafter, we assume $\alpha_s(m_c)/\pi=0.1$.

In order to establish whether the observed $\Delta a_{CP}$ can be accommodated in the
SM or not, it would be important to monitor other observables which are sensitive to
the same (potential) NP effect. In NP scenarios where $\Delta a_{CP}$ mostly arises
from the chromomagnetic operator, the direct CP asymmetries in radiative decays
$D\to P^+ P^-\gamma$ ($P=\pi,K$) are the best candidates to make such a test, as
recently pointed out in ref.~\cite{Isidori:2012yx} (see also ref.~\cite{Lyon:2012fk}).
In particular, the CP violating asymmetries for $D\to V\gamma$ in the $\rho$ and $\omega$
modes can be estimated as~\cite{Isidori:2012yx}
\be
|a_{(\rho,\omega)\gamma}| = 0.04(1) \left|\frac{{\rm Im}[C_7(m_c)]} {0.4\times 10^{-2}} \right|
\left[\frac{10^{-5}}{{\mathcal B}(D\to (\rho,\omega)\gamma)}\right]^{1/2},
\label{eq:maxCPrho}
\ee
where we have assumed maximal strong phases.

Indeed, even if $D\to P^+ P^-\gamma$ is sensitive to $C^{(\prime)}_{7}$, in contrast
to $\Delta a_{CP}$ that is sensitive to $C^{(\prime)}_{8}$, many NP scenarios predict
comparable effects for $C^{(\prime)}_{7,8}$. Moreover, irrespectively of the high scale
value for $C^{(\prime)}_{7}$, a non-vanishing $C^{(\prime)}_{8}$ at the high scale
unavoidably contributes to $C^{(\prime)}_{7}(m_c)$ through QCD running effects.

In the case of SUSY the dominant effects to $C^{(\prime)}_{7,8}$ arise from the loop
exchange of gluinos and up-squarks with an underlying left-right mass insertion
$\left(\delta^{u}_{12}\right)_{LR}$.
Although our numerical analysis is based on the exact expressions for
$C^{(\prime)}_{7,8}$ as evaluated in the mass eigenstate basis~\cite{Misiak},
in the following, for illustrative purposes, we provide the expression for
$C_{7,8}$ at the SUSY scale in the mass-insertion approximation,
\be
\label{eq:C7_g}
C_{7,8}^{(\tilde g)} = -
\frac{\sqrt{2}\pi\alpha_s \tilde m_{g}}{G_F m_c}
\frac{\left(\delta^{u}_{12}\right)_{LR}}{{\tilde m_q^2}}~g_{7,8}(x_{gq})~,
\ee
where $x_{gq}={\tilde m}_g^2/{\tilde m_q^2}$ and the loop functions are
\begin{eqnarray}
g_7^{(2)}(x) &=& \frac{4(1+5x)}{9(1-x)^3} + \frac{8x(2+x)}{9(1-x)^4}\log{x}
~,~~~~~~~~g_7(1)= \frac{2}{27}~,\\
g_8(x) &=& \frac{11+x}{3(1-x)^3} + \frac{9+16x-x^2}{6(1-x)^4}\log{x}~,~~~~g_8(1)
= -\frac{5}{36}\,.
\label{eq:g8}
\end{eqnarray}
In particular, assuming degenerate supersymmetric masses
(${\tilde m_q}=\tilde m_{g}\equiv \tilde m$), one can find
\be
\left|\Delta a^{\rm SUSY}_{CP} \right| \approx 0.6\%
\left(
\frac{
\left|\mathrm{Im} \left( \delta^{u}_{12}\right)_{LR} +
\mathrm{Im} \left( \delta^{u}_{12}\right)_{RL}
\right|}{10^{-3}}
\right)
\left(\frac{{\rm TeV}}{{\tilde m}}\right)~.
\label{eq:acpNPbis}
\ee
On the other hand, we also find that
\be
|a^{\rm SUSY}_{(\rho,\omega)\gamma}| \approx
5\left|\Delta a^{\rm SUSY}_{CP}\right|
\times \left[\frac{10^{-5}}{{\mathcal B}(D\to (\rho,\omega)\gamma)}\right]^{1/2}
\lesssim 10\%~,
\label{eq:CPrho_susy}
\ee
where we have taken $m_{\rm SUSY}=1 \, {\rm TeV}$.

In our setup, the dominant contributions come from $(\delta^u_{LR})_{12}^{eff}$ and
$(\delta^u_{LR})_{21}^{eff}$ as given by Eq.~(\ref{delta2}). Since in the absence of
cancellations the $(\lambda_U)_{i3}$ entries are bounded by the corresponding CKM
elements, the maximal effects are given by
\begin{align}
\label{ACPmax1}
{\rm Im} (\delta^u_{LR})_{12} &\sim 1 \times 10^{-3}   \left( \frac{A}{\tilde{m}} \right)
\left(\frac{1 \, \TeV}{\tilde{m}}\right) \left(\frac{(\lu)_{13}}{\lambda^3}\right) \left(\frac{(\lu)_{32}}{\ord{1}}\right), \\
{\rm Im} (\delta^u_{LR})_{21} &\sim 7 \times 10^{-3}   \left( \frac{ A}{\tilde{m}} \right)
\left(\frac{1 \, \TeV}{\tilde{m}}\right) \left(\frac{(\lu)_{23}}{\lambda^2}\right) \left(\frac{(\lu)_{31}}{\ord{1}}\right),
\label{ACPmax2}
\end{align}
where the parameter $A$ is defined by  $(A_U)_{33} = A\,y_t$. We therefore see that
an imaginary part of the order of $10^{-3}$ can be easily achieved in this setup.

However, non-trivial bounds in the up-sector arise from $D^0-\overline{D}^0$ mixing and the
neutron EDM which constrain $(\delta^u_{RR})_{12}$ and ${\rm Im}(\delta^u_{LR})^{eff}_{11}$,
respectively.
Requiring ${\rm Im}(\delta^u_{RR})_{12} \lesssim 6 \times 10^{-2} (\tilde{m}/\TeV)$ 
(see~\cite{ING, Cetal}
and references therein) and ${\rm Im}(\delta^u_{LR})_{11} \lesssim 4 \times 10^{-6} (\tilde{m}/\TeV)$,\footnote{\label{footnote:EDM} The parameters $(\delta^{d,u}_{LR})_{11}$ are costrained by hadronic EDMs. Imposing the experimental bound from the neutron EDM, $|d_n| < 2.9 \times 10^{-26}~e\,\rm{cm}~(90\% \rm{CL})$~\cite{Baker:2006ts},
we find ${\rm Im}(\delta^{d}_{LR})_{11}\lesssim 2 \times 10^{-6}(\tilde{m}/1~\TeV)$,
${\rm Im}(\delta^{u}_{LR})_{11}\lesssim 4 \times 10^{-6}(\tilde{m}/1~\TeV)$. The neutron EDM
$d_n$ has been estimated in terms of constituent quark EDMs and chromo--EDMs using the result
of ref.~\cite{Hisano:2012sc}, which is based on QCD sum rules~\cite{Pospelov:1999ha,Pospelov:2000bw},
and through the QCD RG evolution from ${\tilde m}$ down to $\sim {\mathcal O}(1)~$GeV~\cite{Degrassi:2005zd}.
Our numerical results are based on the exact expressions for the SUSY contributions
in the mass eigenstate basis of ref.~\cite{Hisano:2008hn}.}
we find
\begin{align}
\im \, [ (\lu^*)_{31}(\lu)_{32}] & \lesssim 6 \times 10^{-2} \left( \frac{\tilde{m}}{1\, \TeV} \right), \label{eq:BOUND1}\\
\im \, [ (\lu)_{13} (\lu)_{31}] & \lesssim 2 \times 10^{-5} \left( \frac{\tilde{m} }{1 \, \TeV} \right) \left( \frac{\tilde{m}}{  A} \right). \label{eq:BOUND2}
\end{align}
Here we have neglected the effect of the different loop function that arises from the effective
double or triple mass insertion. Taking it into account will slightly weaken the bounds.

Since $(\delta^u_{RR})_{12}$ and $(\delta^u_{LR})^{eff}_{11}$ depend on different combinations
of $(\lu)_{ij}$ compared to $(\delta^u_{LR})_{12}$ and $(\delta^u_{LR})_{21}$, it is not possible
to establish model-independently whether the bounds from $D^0-\overline{D}^0$ mixing and EDMs can
be satisfied while simultaneously accounting for $\Delta a^{\rm SUSY}_{CP} \approx 1\%$. 

This can be realized by considering a suitable underlying flavor model that generates a pattern
of Yukawa couplings, which reproduces fermion masses and mixings and respects the constraints
in Eqs.~(\ref{eq:BOUND1}) and (\ref{eq:BOUND2}), while having large $(\lu)_{13} (\lu)_{32}$ or
$(\lu)_{23} (\lu)_{31}$ as in Eqs.~(\ref{ACPmax1}), (\ref{ACPmax2}). For the numerical analysis
in the next section we simply assume that such a model can be constructed. Later on we will
discuss the predictions in the case of a simple $U(1)$ flavor model.

Finally, let us mention that, beside direct CP violation in charm systems, other potentially
interesting observables in this scenario could be rare $B$ and $K$ decays induced by FCNC
$Z$-penguins. However, the leading effects stemming from chargino/up-squark loops are
proportional to $(\delta^{u}_{i3})_{LR}(\delta^{u}_{j3})_{LR} \lesssim V_{i3} V_{j3}$
and therefore too small in order to generate visible effects. On the other hand, the
combination $(\delta^{u}_{3i})_{LR}(\delta^{u}_{3j})_{LR}$ is always accompanied by
light Yukawas and therefore safely negligible.

\subsection{Numerical Results}
\begin{figure}[t]
\centering
\includegraphics[scale=0.5,angle=-90]{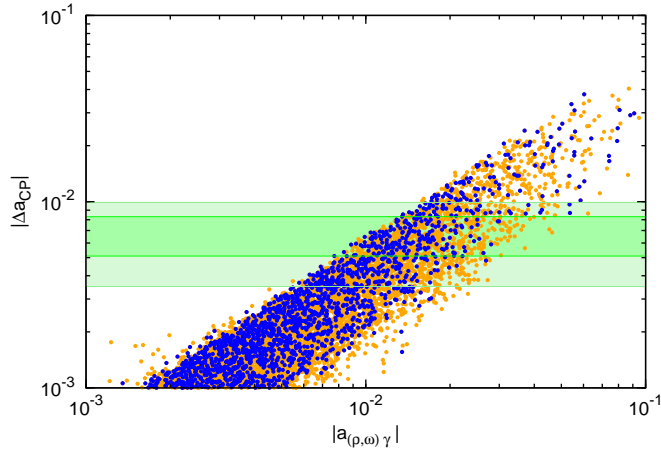}
\caption{\label{fig:DaCP} $|\Delta a_{CP}|$ versus $|a_{(\rho,\omega)\gamma}|$
for a wide scan of the model parameters (see the text for details).
The blue points correspond to $m_h>123$ GeV. The green (dark green)
band represents the 2$\sigma$ (1$\sigma$) experimental range.}
\end{figure}
\begin{figure}[t]
\centering
\includegraphics[scale=0.5,angle=-90]{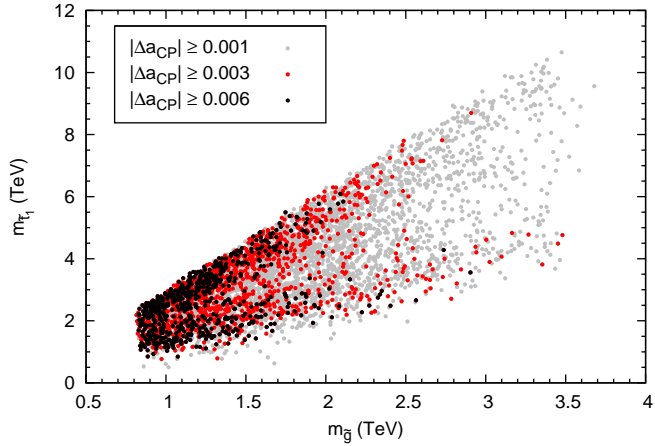}
\caption{\label{fig:glu-stop} Different ranges of $|\Delta a_{CP}|$ displayed in
the $m_{\tilde g}$-$\tilde{m}_{t_1}$ plane. All points correspond to $m_h>123$ GeV.}
\end{figure}
In this section we demonstrate that $(\delta^u_{LR})_{12}^{eff}$ 
and $(\delta^u_{LR})_{21}^{eff}$ as in Eqs.~(\ref{ACPmax1},~\ref{ACPmax2}) can indeed induce
values of $\Delta a_{CP}$ consistent with the experimental observation of Eq.~(\ref{eq:acpExp}).
Furthermore, we illustrate the $\Delta a_{CP}-a_{(\rho,\omega)\gamma}$ correlation of Eq.~(\ref{eq:CPrho_susy}). For the numerical analysis we use the expressions Eqs.~(\ref{eq:Au})-(\ref{eq:m2hd}) for the full structure of soft terms at the high scale.
The gauge mediation parameters are randomly varied in the ranges
\begin{align}
5 \le \tan\beta\le 40, \quad 100~{\rm TeV} \le \Lambda \le 500~{\rm TeV}, \quad %\nonumber \\
\Lambda < M \le 10^{15}~{\rm GeV}, \quad  N=1. \label{eq:flavor-scan}
\end{align}
According to Eqs.~(\ref{ACPmax1},~\ref{ACPmax2}), 
we consider two separate cases for the entries of the matrix $\lu$:
\begin{align}
{\rm (a)}\quad &(\lu)_{ii} = [0.3,3]\times (y_U)_{ii}~~[i=1,2,3], \quad (\lu)_{13} = [0.3,3]\times \lambda^3,
\quad (\lu)_{32} = [0.3,3], \\
{\rm (b)}\quad &(\lu)_{ii} = [0.3,3]\times (y_U)_{ii}~~[i=1,2,3], \quad (\lu)_{23} = [0.3,3]\times \lambda^2,
\quad (\lu)_{31} = [0.3,3].
\end{align}
All the other entries are set to zero and the phases are randomly varied between 0 and 2$\pi$.
In this way we assume an underlying theory of flavor in which the up sector only partially contributes to the CKM matrix. Of course in a realistic model the entries that we neglect here are expected to
be different from zero, but we only need to require that they are small enough in order to satisfy
the constraints in Eqs.~(\ref{eq:BOUND1}) and (\ref{eq:BOUND2}). Concretely, in case (a) we need $(\lu)_{31} < 2\times 10^{-3}$, in case (b) $(\lu)_{32} < 6\times 10^{-2}$ and $(\lu)_{13} < 2\times 10^{-5}$. Moreover, we allow for a moderate hadronic enhancement to the resulting $\Delta a_{CP}$, varying it randomly between 1 and 3. As discussed above, $D^0-\overline{D}^0$ and EDMs constraints
are automatically satisfied for both cases (a) and (b).

The result is shown in Fig.~\ref{fig:DaCP}, where we plot $|\Delta a_{CP}|$ versus $|a_{(\rho,\omega)\gamma}|$, displaying together the points of the scans (a) and (b). 
The green (dark green) band is the 2$\sigma$ (1$\sigma$) experimental range 
reported in Eq.~(\ref{eq:acpExp}). The blue points correspond to $m_h>123$ GeV.

In Fig.~\ref{fig:glu-stop}, different ranges of $|\Delta a_{CP}|$ are displayed in the $m_{\tilde g}$-$\tilde{m}_{t_1}$ plane. All points fulfill the condition $m_h>123$ GeV. The SUSY contribution
to $\Delta a_{CP}$ decouples faster by increasing the gluino mass than the stop mass, as one can
easily check from the behavior of the loop function of $C_8$, see Eq.~(\ref{eq:g8}). Nevertheless,
even for gluinos as heavy as 2$\div$3 TeV one can have $\Delta a_{CP}\gtrsim 6\times 10^{-3}$.

\section{Comparison with other models}
\label{sec:comparison}
In this section we compare the particular flavor structure of FGM to other models that predict 
the parametric flavor suppression of soft terms. In particular, we consider MFV-like models, 
$U(1)$ flavor symmetry models and SUSY models with Partial Compositeness (PC).

In all these models the SUSY mediation scale $\Lambda_S$ is assumed to be above the scale of 
flavor messengers $\Lambda_F$, so that the flavor structure of soft terms at the scale $\Lambda_F$ 
is controlled entirely by the flavor dynamics at this scale, irrespectively of their structure at 
the scale $\Lambda_S$.
In FGM the situation is reversed as the SUSY messenger scale $\Lambda_S = M$ is below $\Lambda_F$.
This setup is therefore complementary to the other scenarios, allowing also for very low SUSY
mediation scales. All the unspecified dynamics of the flavor sector is imprinted in the matter-messenger 
couplings, just like Yukawas, and the full SUSY spectrum is totally calculable in terms of these couplings.

In contrast to the previous section, we will now take a broader perspective for FGM, and consider 
soft terms that have the most general structure expected for superpotentials in which all Yukawa-like 
couplings of light matter fields to the messengers are present (which requires more than one pair of 
messengers). We restrict to the case in which these couplings (along with Yukawas) have the parametric 
suppression expected in $U(1)$ flavor models or almost equivalently Partial Compositeness models. 
Before discussing the general form of soft terms and comparing to the other scenarios, we will briefly 
review the flavor structure of soft terms and mass insertions in the MFV, $U(1)$ and PC cases.

\subsection{Minimal Flavor Violation}

The MFV ansatz is based on the observation that, for vanishing Yukawa couplings,
the SM quark sector exhibits an enhanced global symmetry
\begin{equation}\label{eq:Gflavor}
G_\mathrm{f}~=~SU(3)_u \times SU(3)_d \times SU(3)_Q \,.
\end{equation}
The SM Yukawa couplings are formally invariant under $G_\mathrm{f}$ if the Yukawa matrices
are promoted to spurions transforming appropriately under $G_\mathrm{f}$. New Physics models
are of MFV type if there is no new flavor structure beyond the SM Yukawas. In this case they
are formally invariant under $G_\mathrm{f}$~\cite{MFV}, and the most general
flavor structure can be recovered by a spurion analysis.

In the $R$-parity conserving MSSM, the general expressions for the low-energy soft-breaking
terms compatible with the MFV principle read~\cite{MFV}
\begin{align}
\tilde{m}_{Q}^{2} &\sim \mathbf{1}+ \yu \yud + \yd \ydd,
\label{MFV:mQ}
\end{align}
\vspace{-0.7cm}
\begin{align}
\tilde{m}_{U}^{2} &\sim  \mathbf{1} + \yud \yu +  \yud \yd \ydd \yu, 
&
\tilde{m}_{D}^{2} &\sim \mathbf{1} +  \ydd  \yd + \ydd \yu \yud \yd,
\label{MFV:mUD}
\\
A_{U} &\sim A\left(\mathbf{1} +  \yu \yud  +\yd \ydd  \right) \yu,
&
A_{D} &\sim A\left(\mathbf{1} +  \yu \yud  +\yd \ydd  \right) \yd,
\label{MFV:AUD}
\end{align}
where we omitted $\ord{1}$ complex coefficients and higher-order terms in $y_{U,D}$
(see ref.~\cite{colangelo_MFV} for the most general expressions).
Therefore, keeping only the leading terms, we obtain the following mass insertions
\begin{align}
(\delta_{LL}^u)_{ij} &\sim  V_{i3}^* V_{j3}\, y^2_b,
&
(\delta_{LL}^d)_{ij} & \sim V_{3i}V_{3j}^* \, y^2_t,
\\
(\delta_{RR}^u)_{ij} & \sim y^U_i y^U_j \,V_{i3}^* V_{j3}  \, y_b^2,
&
(\delta_{RR}^d)_{ij}&  \sim y^D_i y^D_j \,V_{3i}V_{3j}^* \, y^2_t,
\\
(\delta_{LR}^u)_{ij} & \sim
\frac{m^U_j A}{{\tilde m}_Q{\tilde m}_U} \,V_{i3}^* V_{j3} \, y^2_b,
&
(\delta_{LR}^d)_{ij} & \sim 
\frac{m^D_j A}{{\tilde m}_Q{\tilde m}_D} \,V_{3i}V_{3j}^* \, y^2_t.
\end{align}
Notice that the dominant flavor violating effects typically arise from the LL and LR
down sectors due to the large top Yukawa coupling, unless $y_b \sim 1$ corresponding
to large $\tan\beta \sim m_t/m_b$.
Moreover, the symmetry principle of MFV allows for the presence
of flavor-blind CPV phases~\cite{Kagan:2009bn} and therefore the $\mu$-term, the gaugino
masses $M_i$ and trilinear scalar couplings $A_{U(D)}$ might be 
complex~\cite{colangelo_MFV, Bartl:2001wc, Ellis:2007kb, Altmannshofer:2008hc}.
This, in turn, leads to exceedingly large CPV phenomena such as the neutron EDM (through
the one loop exchange of gauginos and sfermions), unless the first generation sfermions
are heavy~\cite{Altmannshofer:2008hc, Paradisi:2009ey, Barbieri:2011vn} or some other
mechanism is at work to suppress these CPV phases~\cite{Paradisi:2009ey}.

\subsection{$U(1)$ Flavor Models}
In $U(1)$ flavor symmetry models Yukawas are of the (hierarchical) form
\begin{align}
\label{U1model}
(y_U)_{ij}  & \sim \eps^{Q_i + U_j}, & (y_D)_{ij}  & \sim \eps^{Q_i + D_j},
\end{align}
where $\eps$ is a small order parameter and $Q_i,U_i, D_i$ denote the positive
$U(1)$ charges of the respective superfields. Using $Q_3 = U_3 = 0$ as suggested
by the large top Yukawa, all other charges can be expressed in terms of diagonal
Yukawa couplings and CKM matrix elements, giving
\begin{align}
\eps^{Q_i} & \sim V_{i3}, & \eps^{U_i} & \sim \frac{y^U_i}{V_{i3}}, & \eps^{D_i} & \sim \frac{y^D_i}{V_{i3}}.
\label{eq:u1}
\end{align}
The structure of the soft masses as determined by $U(1)$ invariance is given by
\begin{align}
\tilde{m}_{Q}^{2} & \sim  \eps^{|Q_i - Q_j|},
&
\tilde{m}_{U}^{2} &\sim  \eps^{|U_i - U_j|}, 
&
\tilde{m}_{D}^{2} & \sim \eps^{|D_i - D_j|},
\end{align}
\begin{align}
A_{U} &\sim \eps^{Q_i +U_j},
&
A_{D} &\sim \eps^{Q_i +D_j},
\end{align}
so that one obtains for the ``mass insertions'' (MIs)
\begin{align}
(\delta_{LL}^u)_{ij} &\sim  \frac{V_{i3}}{V_{j3}} |_{i \le j},
&
(\delta_{LL}^d)_{ij} & \sim  \frac{V_{i3}}{V_{j3}} |_{i \le j},
\\
(\delta_{RR}^u)_{ij} & \sim  \frac{y^U_i V_{j3}}{y^U_j V_{i3}} |_{i \le j},
&
(\delta_{RR}^d)_{ij} & \sim \frac{y^D_i V_{j3}}{y^D_j V_{i3}} |_{i \le j},
\\
(\delta_{LR}^u)_{ij} & \sim \frac{m^U_j A}{{\tilde m}_Q{\tilde m}_U}\frac{V_{i3}^*}{V_{j3}},
&
(\delta_{LR}^d)_{ij} & \sim \frac{m^D_j A}{{\tilde m}_Q{\tilde m}_D} \frac{V_{i3}^*}{V_{j3}},
\end{align}
where in LL and RR the $i>j$ entries are obtained by hermitian conjugation, and in LR we
introduced a complex conjugation to indicate that the diagonal entries are in general complex.

The major problem with the above flavor structures is to satisfy the constraint
from $\epsilon_K \sim (\delta_{LL}^d)_{12}(\delta_{RR}^d)_{12}\sim m_d/m_s$
which typically requires a SUSY scale of $\ord{100} \TeV$. To less extent, also
$\epsilon^{\prime}/\epsilon \sim (\delta_{LR}^d)_{12(21)}$ and the neutron EDM
(which is dominantly generated by the down-quark EDM) provide strong bounds on
$U(1)$ flavor models. Moreover, similarly to the MFV case, the flavor $U(1)$ symmetry
does not prevent the existence of flavor-blind CPV phases for the gaugino masses,
trilinear terms and the $\mu$-term. Therefore, the SUSY CP problem has to be
addressed by some other protection mechanisms in order to make this scenario viable.

%%%%%%%%%%%%%%%%%%%%%%%%%%%%%%%%%%%%%%%%%%%%%%%%%%%%%%%%
\subsection{Partial Compositeness}  \label{sec:PC}
%%%%%%%%%%%%%%%%%%%%%%%%%%%%%%%%%%%%%%%%%%%%%%%%%%%%%%%%

Partial Compositeness (PC) is a seesaw-like mechanism that explains the hierarchy among
the SM fermion masses by mixing with heavy resonances of a strongly coupled sector.
Originally proposed within Technicolor models~\cite{Kaplan:1991dc}, it has been
subsequently applied to extra-dimensional RS models~\cite{Gherghetta:2000qt,Agashe:2004cp} 
and also in the context of SUSY~\cite{Nomura, Lodone}.

The basic assumption is that at the UV cutoff the SM fermions couple linearly to operators of
the strong sector that is characterized  by the mass scale $m_\rho$ and the coupling $g_\rho$.
According to the paradigm of Partial Compositeness, in the effective theory below the scale 
$m_\rho$ of the heavy resonances, every light quark $(q, u, d)_i$ is accompanied by a spurion $\epsilon^{q,u,d}_i\lesssim1$ that measures its amount of compositeness. The quark Yukawa
matrices then take the form
\begin{equation}
\label{eq:yukawas}
(y_U)_{ij} \sim g_{\rho} \epsilon^q _i \epsilon^u_j,~~~~~~~~~~~~~~~~~(y_D)_{ij} \sim g_{\rho} \epsilon^q _i \epsilon^d_j,
\end{equation}
which closely resembles the case of a single $U(1)$ flavor model, see  Eq.~(\ref{U1model}),
with the correspondence
\be
\epsilon^{q,u,d} _i \,  \longleftrightarrow \,
\epsilon^{Q_i,U_i,D_i}.
 \ee
A slight difference arises from the presence of the coupling $g_\rho$ that can be large in this case. This implies that one can consider also $\eps^q_3, \eps^u_3 < 1$ or equivalently $Q_3, U_3 \ne 0$, since the top Yukawa can arise from strong coupling. One has therefore two more parameters that we choose as $\eps^q_3$ and $\eps^u_3$. 

Apart from this issue, there is no difference between a single $U(1)$ and PC for what regards Yukawa couplings, or in general all superpotential terms. The main difference is in the non-holomorphic soft terms, which at the scale $m_{\rho}$ are expected to be of the form~\cite{Lodone}
\begin{align}
\tilde{m}_{Q}^{2} &\sim \mathbf{1} + \eps^q_i \eps^q_j,
\end{align}
\vspace{-0.7cm}
\begin{align}
\tilde{m}_{U}^{2} &\sim  \mathbf{1} +   \eps^u_i \eps^u_j, 
&
\tilde{m}_{D}^{2} &\sim \mathbf{1} +   \eps^d_i \eps^d_j,
\\
A_{U} &\sim g_\rho \eps^q_i \eps^u_j,
&
A_{D} &\sim  g_\rho \eps^q_i \eps^d_j,
% \label{MFV:AUD}
\end{align}
Therefore we find the following MIs
\begin{align}
(\delta_{LL}^u)_{ij} &\sim  (\eps^q_3)^2 V_{i3}^* V_{j3},
&
(\delta_{LL}^d)_{ij} & \sim  (\eps^q_3)^2 V_{3i}V_{3j}^*,
\\
(\delta_{RR}^u)_{ij} & \sim  \frac{y^U_i y^U_j} {V_{i3}^* V_{j3}} \frac{(\eps^u_3)^2}{\yt^2} ,
&
(\delta_{RR}^d)_{ij} & \sim \frac{y^D_i y^D_j} {V_{3i} V_{3j}^*} \frac{(\eps^u_3)^2}{\yt^2} ,\\
(\delta_{LR}^u)_{ij} & \sim  \frac{m^U_j A}{{\tilde m}_Q{\tilde m}_U}\frac{V_{i3}^*}{V_{j3}},
&
(\delta_{LR}^d)_{ij} & \sim \frac{m^D_j A}{{\tilde m}_Q{\tilde m}_D}\frac{V_{3i}}{V_{3j}^*}.
\end{align}
Comparing the flavor structure for the soft sector in the $U(1)$ model and PC cases,
the most prominent feature is the higher suppression for off-diagonal sfermion masses
in the LL and RR sectors in the PC case. The LR sector has the same parametric structure
in PC and $U(1)$ models, since in both scenarios the A-terms are proportional to the
SM Yukawas. Moreover, PC and $U(1)$ flavor models share also the same SUSY CP problem,
as the PC paradigm does not protect against flavor-blind CPV phases.

\subsection{Flavored Gauge Mediation}
In order to be general, we now consider the case in which all Yukawa-like couplings of light
matter fields to messenger fields are present in the superpotential. As discussed, this
requires the presence of more than one messenger pair, with a messenger $(\Phi_{H_u})_1$ that
has the same quantum number as $H_u$ and a messenger $(\overline{\Phi}_{H_d})_2$ that has the
same quantum number as $H_d$, for what regards the symmetry that forbids the $\mu$-term.
The new superpotential terms are then of the form
\be
\Delta W = (\lu)_{ij} Q_i U_j (\Phi_{H_u})_1 +  (\lambda_d)_{ij} Q_i D_j (\overline{\Phi}_{H_d})_2.
\ee
The flavor structure of the soft terms can be found by a spurion analysis noting that the new
couplings transform under the global flavor group as the corresponding Yukawas. The result is
\begin{align}
A_U & \sim \ld \ldd \yu + \lu \lud \yu + \yu \lud \lu, & A_D & \sim  \ld \ldd \yd + \lu \lud \yd + \yd \ldd \ld,
\end{align}
\begin{align}
\Delta \tilde{m}^2_Q & \sim \lu \lud, & \Delta \tilde{m}^2_U & \sim \lud \lu, & \Delta \tilde{m}^2_D  \sim  \ldd \ld. &
\end{align}
At this point we restrict to the case where a single $U(1)$ symmetry controls the size
of the superpotential couplings, that is we take
\begin{align}
(\yu)_{ij} & \sim
(\lu)_{ij} \sim \eps^{Q_i + U_j}, & (\yd)_{ij} & \sim (\ld)_{ij} \sim \eps^{Q_i + D_j}.
\end{align}
The mass insertions can then be calculated in terms of Yukawas and CKM elements.
Keeping only the leading-order terms, one obtains
\begin{align}
\label{deltaLL}
(\delta^u_{LL})_{ij}  & \sim V_{i3}^* V_{j3} \yt^2, & (\delta^d_{LL})_{ij} & \sim  V_{3i} V_{3j}^* \yt^2, \\
\label{deltaRR}
(\delta^u_{RR})_{ij} & \sim \frac{y^U_i y^U_j}{V_{i3}^* V_{j3}}, &
(\delta^d_{RR})_{ij} & \sim  \frac{y^D_i y^D_j}{V_{3i} V_{3j}^*},
\end{align}
\begin{equation}
(\delta^u_{LR})_{ij} \sim \frac{m^U_j A}{{\tilde m}_Q{\tilde m}_U}
\left( V_{i3}^* V_{j3} + \frac{y^U_i y^U_i}{V_{i3}^* V_{j3} \yt^2 } \right),
\end{equation}
\begin{equation}
(\delta^d_{LR})_{ij} \sim  \frac{m^D_j A}{{\tilde m}_Q{\tilde m}_D}
\left( V_{3i} V_{3j}^* + \frac{y^D_i y^D_i}{V_{3i} V_{3j}^* \yt^2 } \right).
\end{equation}
For light generations the dominant effective MIs arise from the contractions $ \delta^{q}_{LL}\delta^{q}_{LR}\delta^{q}_{RR}$, $ \delta^{q}_{LL}\delta^{q}_{LR}$ or $ \delta^{q}_{LR}\delta^{q}_{RR}$. For the leading-order terms we find
\begin{equation}
(\delta^u_{LR})_{ij}^{eff} \sim
\frac{m^U_j \left(A - \yt^2 \mu^*/\tan\beta\right)}{{\tilde m}_Q{\tilde m}_U}
\frac{V_{i3}^*}{V_{j3}} \yt^2~\qquad i,j = 1,2,
\end{equation}
\begin{equation}
(\delta^d_{LR})_{ij}^{eff} \sim
\frac{m^D_j \left(A - \yt^2 \mu^* \tan\beta\right)}{{\tilde m}_Q{\tilde m}_D}
\frac{V_{3i}} {V_{3j}^*} \yb^2~\qquad i,j = 1,2.
\end{equation}
Note that in contrast to the other models $(\delta^u_{LR})_{12}^{eff}$ is larger than $(\delta^u_{LR})_{12}$, because the up--Yukuwa in $A_U$ can be sandwiched between $\lu$
spurions, avoiding the double suppression by light Yukawas and CKM factors. The same is
true also for $(\delta^d_{LR})_{12}^{eff}$ provided that $\tan \beta$ is large enough.
Finally, for $ij=3j,i3$, it turns out that $(\delta^u_{LR})_{ij}^{eff} \sim (\delta^u_{LR})_{ij}$,
while $(\delta^d_{LR})_{ij}^{eff}$ can be larger than $(\delta^d_{LR})_{ij}$ if
$(\mu/A) \tan\beta > 1$.

Despite the underlying $U(1)$ flavor symmetry, the GMSB setup leads to a strong suppression
for the soft terms that rather resembles the PC structure. Here the extra suppression 
originates in the loop origin of soft terms, which acts precisely as a wavefunction suppression \cite{Pokorski, Isidori}.
Indeed LL and RR mass insertions are suppressed as in PC, while (effective) LR mass insertions in
the up sector have roughly the same suppression. Interestingly LR in the down sector is additionally suppressed by $y_b^2$, which becomes strong in the low $\tan \beta$ regime. This setup therefore
allows the implementation of SUSY flavor models with a built-in suppression of $\Delta F=2$ effects
and flavor-violating effects mainly arising from the LR mass insertions.

Moreover, as opposite to the case of MFV, PC and $U(1)$ flavor models, the FGM setup
has also a built-in protection against flavor-blind CPV phases stemming from GMSB. Yet, the 
$\mu$- and $B_\mu$-terms, which are not controlled by GMSB, could introduce irremovable phases 
depending on the underlying mechanism that generates them. Still, if this mechanism is such 
that the phases of $\mu$ and $B_\mu$ are correlated to the phase of $\Lambda$, then no phases 
arises from this sector~\cite{GMSB}.

\subsection{Comparison}
We now compare the parametric flavor suppression of the soft terms in the various
scenarios summarized in Table~\ref{table}. In addition to the general FGM discussed 
in the previous section, denoted as FGM$_{U,D}$, we include the model from Section 5 
(where only $\Phi_{H_u}$ couples to matter fields), which in the following we denote 
as FGM$_{U}$.
\begin{table}[t]
\centering
\begin{tabular}{|c||c|c|c|c|c|}
\hline
& MFV &  PC &  $U(1)$ &  FGM$_{U,D}$ \!+\! $U(1)$ &  FGM$_U$ \!+\! $U(1)$ \\
\hline
& & & & & \\
$(\delta^u_{LL})_{ij} $ & $ V_{i3}^* V_{j3} \yb^2$ & $   V_{i3}^* V_{j3} (\eps^q_3)^2 $ & $\frac{V_{i3}}{V_{j3}} |_{i \le j}$ & $V_{i3} ^*V_{j3} $ & $V_{i3}^* V_{j3} $ \\
& & & & & \\
$(\delta^d_{LL})_{ij} $ & $V_{3i} V_{3j}^* $ & $V_{3i} V_{3j}^* (\eps^q_3)^2   $ & $\frac{V_{i3}}{V_{j3}} |_{i \le j}$  & $V_{3i} V_{3j}^* $ & $V_{3i} V_{3j}^* $ \\
& & & & & \\
$(\delta_{RR}^u)_{ij} $ & $y^U_i y^U_j V_{i3}^* V_{j3} \yb^2$& $ \frac{y^U_i y^U_j}{V_{i3}^* V_{j3}} (\eps^u_3)^2 $  & $\frac{y^U_i V_{j3}}{y^U_j V_{i3}} |_{i \le j} $ & $\frac{y^U_i y^U_j}{V_{i3}^* V_{j3}}$  & $\frac{y^U_i y^U_j}{V_{i3}^* V_{j3}}$ \\
& & & & & \\
$(\delta_{RR}^d)_{ij} $ & $y^D_i y^D_j V_{3i} V_{3j}^* $&  $ \frac{y^D_i y^D_j}{V_{3i} V_{3j}^*}  (\eps^u_3)^2 $&  $\frac{y^D_i V_{j3}}{y^D_j V_{i3}} |_{i \le j} $&$\frac{y^D_i y^D_j}{V_{3i} V_{3j}^*}$ & $y^D_i y^D_j V_{3i} V_{3j}^*  $ \\
& & & & & \\
\multirow{2}{1cm}{\centering $(\delta_{LR}^u)_{ij} $ }& $y^U_j V_{i3}^* V_{j3}   \yb^2 $& $ y^U_j \frac{V_{i3}^*}{V_{j3}}  $&  $y^U_j  \frac{V_{i3}^*}{V_{j3}}$& 
$ y^U_j ( V_{i3}^* V_{j3} \!+\! \frac{y^U_i y^U_i}{V_{i3}^* V_{j3}   } ) $ & 
$ y^U_j ( V_{i3}^* V_{j3} \!+\! \frac{y^U_i y^U_i}{V_{i3}^* V_{j3}   } ) $ \\
 & & & & $y^U_j \frac{V_{i3}^*}{V_{j3}} $ &  $y^U_j \frac{V_{i3}^*}{V_{j3}} $  \\
& & & & &\\
\multirow{2}{1cm}{\centering $(\delta_{LR}^d)_{ij} $ } & $y^D_j  V_{3i} V_{3j}^*  $ & $ y^D_j \frac{V_{3i}}{V_{3j}^*} $ &
$y^D_j \frac{V_{i3}^*}{V_{j3}}$& 
$y^D_j ( V_{3i} V_{3j}^*  \!+\!  \frac{y^D_i y^D_i}{V_{3i} V_{3j}^*}$ ) &
$ y^D_j  V_{3i} V_{3j}^* $\\
& & & & $y^D_j \frac{V_{3i}}{V_{3j}^*}  y_b^2$ &\\
\hline 
\end{tabular}
\caption{\label{table} Parametric suppression for mass insertions in various scenarios.
The entries in the $U(1)$ column with $i>j$ are obtained from hermiticity. In the LR rows
for FGM we included the effective mass insertions $\delta_{LR}^{eff}$ proportional to $A$
in the lower entry when they can dominate over the direct ones in the upper entry.
We neglect powers of $\yt$.}
\end{table}
\begin{table}[t]
\centering
\begin{tabular}{|c||c|c|c|c|c|c|c|}
\hline
& MFV &  PC &  $U(1)$ &  FGM$_{U,D}$ \!+\! $U(1)$ &  FGM$_U$ \!+\! $U(1)$ & EXP. & OBS. \\
\hline
& & & & & & &\\
$\langle \delta^d\rangle^{2}_{12}$ &
$ y_d y_s \lambda^{10} $ &
$\frac{y_d y_s}{ g^2_\rho}$ &
$\frac{y_d}{y_s}$ &
$y_d y_s $ &
$y_d y_s \lambda^{10} $ &
$7 \!\times\! 10^{-8}$ &
$\epsilon_K$\\
& & & & & & &\\
$\langle \delta^u\rangle^{2}_{12}$ &
$y_u y_c \lambda^{10} \yb^4$ &
$\frac{y_u y_c}{ g^2_\rho}$ &
$\frac{y_u}{y_c}$ &
$y_u y_c $  &
$y_u y_c $ &
$ 1 \!\times\! 10^{-5}$ &
$|q/p|$, $\phi_D$
\\
& & & & & & &\\
$(\delta_{LR}^u)_{12} $ &
$\frac{m_{c} a_U}{{\tilde m}^2} \lambda^5 \yb^2$ &
$\frac{m_{c} A}{{\tilde m}^2} \lambda$ &
$\frac{m_{c} a_U}{{\tilde m}^2} \lambda$ &
$ \frac{m_{c} a_U}{{\tilde m}^2} \lambda $ &
$ \frac{m_{c} a_U}{{\tilde m}^2} \lambda $ &
$2 \!\times\! 10^{-3}$ &
$\Delta a_{CP}$
\\
& & & & & & &\\
$(\delta_{LR}^d)_{12}$ &
$\frac{m_{s} a_D}{{\tilde m}^2} \lambda^5  $ &
$\frac{m_{s} A}{{\tilde m}^2} \lambda$ &
$\frac{m_{s} a_D}{{\tilde m}^2} \lambda$ &
$ \frac{m_{s} a_D}{{\tilde m}^2} \lambda  \yb^2$ &
$ \frac{m_{s} a_D}{{\tilde m}^2} \lambda^5  $ &
$ 4 \!\times\! 10^{-5}$ &
$\epsilon^{\prime}/\epsilon$
\\
& & & & & & &\\
$(\delta_{LR}^{u})_{11} $ &
$\frac{m_{u} a_U}{{\tilde m}^2}$ &
$\frac{m_{u} a_U}{{\tilde m}^2}$ &
$\frac{m_{u} a_U}{{\tilde m}^2}$ &
$ \frac{m_{u} a_U}{{\tilde m}^2} $ &
$ \frac{m_{u} a_U}{{\tilde m}^2} $ &
$ 4 \!\times\! 10^{-6}$ &
$d_n$
\\
& & & & & & &\\
$(\delta_{LR}^{d})_{11} $ &
$\frac{m_{d} a_D}{{\tilde m}^2}$ &
$\frac{m_{d} a_D}{{\tilde m}^2}$ &
$\frac{m_{d} a_D}{{\tilde m}^2}$ &
$ \frac{m_{d} a_D}{{\tilde m}^2} \yb^2$ &
$ < \frac{m_{d} A}{{\tilde m}^2} \lambda^6 $ &
$ 2 \!\times\! 10^{-6}$ &
$d_n$
\\
& & & & & & &\\
\hline
\end{tabular}
\caption{\label{table2}
Predictions for the relevant mass insertions in the scenarios of Table~\ref{table}.
Here $\langle\delta^q\rangle^{2}_{12}\equiv(\delta^q_{LL})_{12}(\delta^q_{RR})_{12}$,
$\lambda \approx 0.2$ and we neglect powers of $\yt$. We denote
$a_U \equiv A - \mu^*/\tan \beta $ and  $a_D \equiv A - \mu^*\tan \beta $, where the
parameter $A$ is defined by $(A_{U})_{33} = A \, y_t$ in all scenarios.
The experimental bounds (EXP.) refer to the imaginary components of the MIs for
${\tilde m}= 1~$TeV and are obtained imposing the experimental constraints on the
most relevant processes listed in the last column (OBS.).
}
\end{table}
As can be seen from this table, the MFV scenario always yields the strongest
suppressions and $U(1)$ the weakest. A closer look at the $LL/RR$ and $LR$
MIs of Table~\ref{table} leads to the following general conclusions:

\begin{description}
\item[{\it LL/RR mixing:}] The $U(1)$ model has a much milder suppression compared
to the PC and general FGM cases. In particular, assuming the approximate relation
$y_i/y_j \sim (V_{i3}/V_{j3})^2$, it turns out that
$(\delta^{u,d}_{AA})_{ij} \sim V_{i3}/V_{j3}$ (for $i<j$ and $AA=LL,RR$) in the $U(1)$
case, while $(\delta^{u,d}_{AA})_{ij} \sim V_{i3}V_{j3}$ in the PC and general FGM cases.
This higher suppression is reminiscent of what happens in the case of wave function renormalization~\cite{Pokorski, Isidori} where the LL and RR MIs depend on the sum
of charges instead of their difference in contrast to $U(1)$ models.
In the PC case, we have a further suppression of order $(\epsilon^{q,u}_3)^2$ for
$\delta^{u,d}_{LL,RR}$ compared to FGM$_{U,D}$ which is maximized for maximal
strong couplings $g_\rho \sim 4\pi$ as the top mass relation implies that
$g_\rho \epsilon^q_3\epsilon^u_3 = 1$ with $\epsilon^{q,u}_{3}<1$. While the up sector
is identical in both FGM models, the down sector of FGM$_U$ is MFV-like because there
is no new spurion transforming under $SU(3)_d$. 
\item[{\it LR mixing:}] PC has the same suppression as $U(1)$ in both the up and down sectors.
The FGM$_{U,D}$ shares this suppression in the (effective) LR up-sector, while the
LR down-sector involves an additional $\yb^2$. Again the down sector of FGM$_{U}$ has an additional
suppression that becomes as strong as in MFV.
\end{description}

We now analyze the phenomenological implications of the flavor structure of sfermion
masses in low-energy processes. In particular, we will distinguish among $\Delta F =2$,
$\Delta F =1$, and $\Delta F =0$ processes, where in the latter case we refer to
flavor conserving transitions like the EDMs that are still sensitive to flavor
effects. Concerning $\Delta F = 2,1$ transitions, we will focus only on processes
with an underlying $s\to d$ or $c\to u$ transition as they put the most stringent
bounds to the model in question. The predictions for the most relevant combinations
of MIs are summarized in Table~\ref{table2}.

\begin{description}
\item[{\it $\Delta F =2$ processes:}] the relevant processes here are $K^0-{\bar K}^0$ and 
$D^0-{\bar D}^0$ mixings. As it is well known, these processes are mostly sensitive to the 
combinations of MIs $(\delta^{d}_{LL})_{12}(\delta^{d}_{RR})_{12}$ and $(\delta^{u}_{LL})_{12}(\delta^{u}_{RR})_{12}$,
respectively. In the $U(1)$ case, it turns out that $(\delta^{d}_{LL})_{12}(\delta^{d}_{RR})_{12}\sim m_d/m_s \approx 0.05$,
which implies a very heavy SUSY spectrum given the model-independent bound from $\epsilon_K$
that requires
${\rm Im}[(\delta^{d}_{LL})_{12}(\delta^{d}_{RR})_{12}]\lesssim 10^{-7}~(\tilde{m}/1\,\TeV)$.
The $D^0-{\bar D}^0$ bounds are automatically satisfied after imposing that from $\epsilon_K$.
The situation greatly improves in the PC and FGM$_{U,D}$ cases where we have
$(\delta^{d}_{LL})_{12}(\delta^{d}_{RR})_{12}\sim\frac{m_d m_s}{g^{2}_{\rho}v^2}\tan^2\beta
\approx 5 \times 10^{-9}\frac{\tan^2\beta}{g^{2}_{\rho}}$ and $(\delta^{d}_{LL})_{12}(\delta^{d}_{RR})_{12}\sim (m_dm_s/v^2)\times \tan^2\beta
\approx 5 \times 10^{-9}\tan^2\beta$, respectively.
For moderate/small values of $\tan\beta$ and considering ${\mathcal O}(1)$ unknowns,
both PC and FGM$_{U,D}$ scenarios are viable for TeV scale soft masses.
Yet, in the PC and FGM$_{U,D}$ models, it is easy to generate sizable
NP effects for $\epsilon_K$ (but not for $B_{d,s}$ mixing) which can improve
the UT fit~\cite{UTfit}.
Finally MFV and FGM$_{U}$ have an additional CKM suppression in the down sector, which
completely removes the bounds from $\eps_K$ even for large $\tan \beta$. Although the
up sector in FGM$_{U}$ is not as much suppressed as MFV, it is still small enough to
easily satisfy the $D^0-{\bar D}^0$ bounds.
\item[{\it $\Delta F =1$ processes:}] The most constraining process of this
sector is $\epsilon^{\prime}/\epsilon$, which provides the model-independent bound
${\rm Im}(\delta^{d}_{LR})_{12}\lesssim 4 \times 10^{-5}(\tilde{m}/1~\TeV)$.
Such an upper bound can be saturated in PC and $U(1)$ models where
$(\delta^{d}_{LR})_{12}\sim (A/\tilde{m})\times(m_s\lambda/\tilde{m})$.
Imposing the vacuum stability condition $A/\tilde{m}\lesssim 3$, it turns
out that $(\delta^{d}_{LR})_{12}\lesssim 4\times 10^{-5}(1~\TeV/\tilde{m})$.
In the FGM$_{U,D}$ case, there is an additional $\yb^2$ suppression in the LR
down-sector compared to the PC and $U(1)$ cases and therefore $\epsilon^{\prime}/\epsilon$
does not put any constraint to the model especially for moderate/low $\tan\beta$ values.
The situation further improves in the FGM$_U$ scenario, in which the down LR sector
becomes as strong as in MFV.
\item[{\it $\Delta F =0$ processes:}]
Hadronic EDMs constrain the MIs $(\delta^{d,u}_{LR})_{11}$. Imposing the experimental bound
on the neutron EDM~\cite{Baker:2006ts}, we find that ${\rm Im}(\delta^{d}_{LR})_{11}\lesssim
2 \times 10^{-6}(\tilde{m}/1~\TeV)$ and ${\rm Im}(\delta^{u}_{LR})_{11}\lesssim 4 \times 10^{-6}(\tilde{m}/1~\TeV)$ (see footnote~\ref{footnote:EDM}). In $U(1)$ and PC models,
assuming $A/\tilde{m}\simeq 3$ and the PDG values for $m_{u,d}$~\cite{PDG}, it turns
out that $(\delta^{d}_{LR})_{11}\sim 8 \times 10^{-6}(1~\TeV/\tilde{m})$ and
$(\delta^{u}_{LR})_{11}\sim 4 \times 10^{-6}(1~\TeV/\tilde{m})$, which are somewhat
in tension with the hadronic EDM bounds especially in the down sector.
In the FGM$_{U,D}$ case, there is an additional $\yb^2$ suppression in the LR down-sector
and therefore the EDM constraints are significantly relaxed for small and moderate $\tan\beta$.
The LR-down sector of FGM$_U$ has an additional CKM suppression that is even stronger
than in MFV and removes the down EDM constraint completely.
\end{description}

We now turn to the predictions for $\Delta a_{CP}$ in the models of Table~\ref{table}.
First of all, in MFV the effect is way too small to explain the observed value, as
$(\delta_{LR}^u)_{12}\sim {\mathcal O}(10^{-7})$ even for $y_b \sim 1$. Instead $U(1)$
and PC scenarios have already been used to address the CP asymmetry within a SUSY context
in Refs.~\cite{Nir} and \cite{Lodone}, respectively.

The conclusions in the $U(1)$ case were not too optimistic, mainly because of the tight
constraints from $\epsilon_K$ that can be relaxed only by relying on large gluino RG 
contributions to provide some degree of degeneracy of sfermions~\cite{Dine:1990jd, Nir}.
Moreover, the $\epsilon^{\prime}/\epsilon$ constraint, which requires
$(\delta^d_{LR})_{12} \lesssim 4 \times 10^{-5} (\tilde{m}/\TeV)$, implies the bound
\be
(\delta^u_{LR})_{12} \sim \frac{m_c}{m_s} (\delta^d_{LR})_{12} \lesssim
5 \times 10^{-4} \frac{\tilde{m}}{\TeV},
\ee
which is only marginally compatible with $(\delta^u_{LR})_{12}\sim 10^{-3}$,
as needed for the observed $\Delta a_{CP}$. Finally, the neutron EDM bound implies
the following constraints
\begin{eqnarray}
(\delta^u_{LR})_{12} &\sim& \frac{m_c}{m_d } V_{us} (\delta^d_{LR})_{11}
\lesssim \frac{m_c}{m_d } V_{us} \left( 2 \times 10^{-6} \frac{\tilde{m}}{\TeV} \right)
\sim 1 \times 10^{-4} \frac{\tilde{m}}{\TeV}~,
\label{eq:dlr12u_vs_dlr11d}
\\
(\delta^u_{LR})_{12} &\sim& \frac{m_c}{m_u } V_{us} (\delta^u_{LR})_{11}
\lesssim \frac{m_c}{m_u } V_{us} \left( 4 \times 10^{-6} \frac{\tilde{m}}{\TeV} \right)
\sim 4  \times 10^{-4} \frac{\tilde{m}}{\TeV}~,
\label{eq:dlr12u_vs_dlr11u}
\end{eqnarray}
which have been obtained switching on the down- and up-quark chromo--EDM contributions
to $d_n$ at a time, respectively. As a result, also the EDM bounds are challenging
the explanation of the observed $\Delta a_{CP}$ within $U(1)$ models.
Yet, given the large uncertainties in the evaluation of hadronic EDMs and taking into
account also $\ord{1}$ coefficients, which can easily lead to an accidental enhancement,
one cannot conclude that the EDM constraints prevent the explanation of the observed
effect in $\Delta a_{CP}$.

This situation can be improved in models with Partial Compositeness~\cite{Lodone}
as the constraint from $\epsilon_K$ can be easily satisfied. Yet, the constraints
from the EDMs and $\epsilon^{\prime}/\epsilon$, which are exactly the same as in
the $U(1)$ model, represent a serious challenge for PC models when attempting to
explain $\Delta a_{CP}$.

Passing to FGM$_{U,D}$, we observe that the $\epsilon_K$ constraint can be
quite easily satisfied as in the PC case. Moreover, the $\epsilon^{\prime}/\epsilon$
and EDM bounds are significantly relaxed in this case thanks to the additional $\yb^2$
suppression in LR down-sector.  The situation even improves in the FGM$_U$ scenario, in which the down LR sector becomes MFV-like. Notice that importantly the up LR sector in both FGM setups remains as large as in $U(1)$ and PC. Therefore only the up-quark EDM puts slight constraints on the viable parameter space. We also want to emphasize that in contrast to the PC case, where the presence of
the strongly interacting sector leads to a lack of predictivity even for the flavor-diagonal
SUSY spectrum, the FGM scenario has the main advantage that the SUSY spectrum is similarly
predictive as minimal Gauge Mediation with only one additional parameter.
\begin{figure}[t]
\centering
\includegraphics[scale=0.4,angle=-90]{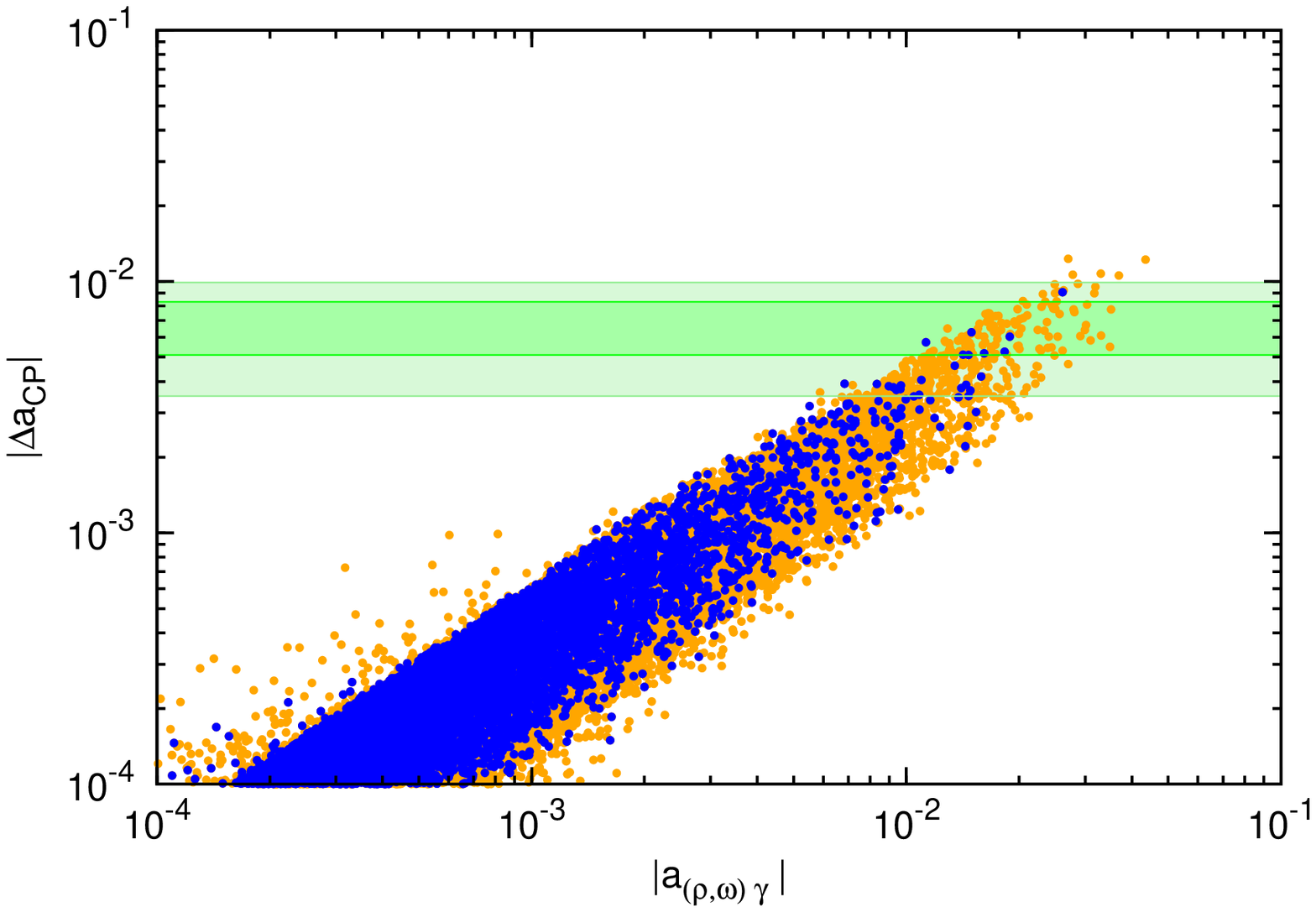}
\includegraphics[scale=0.4,angle=-90]{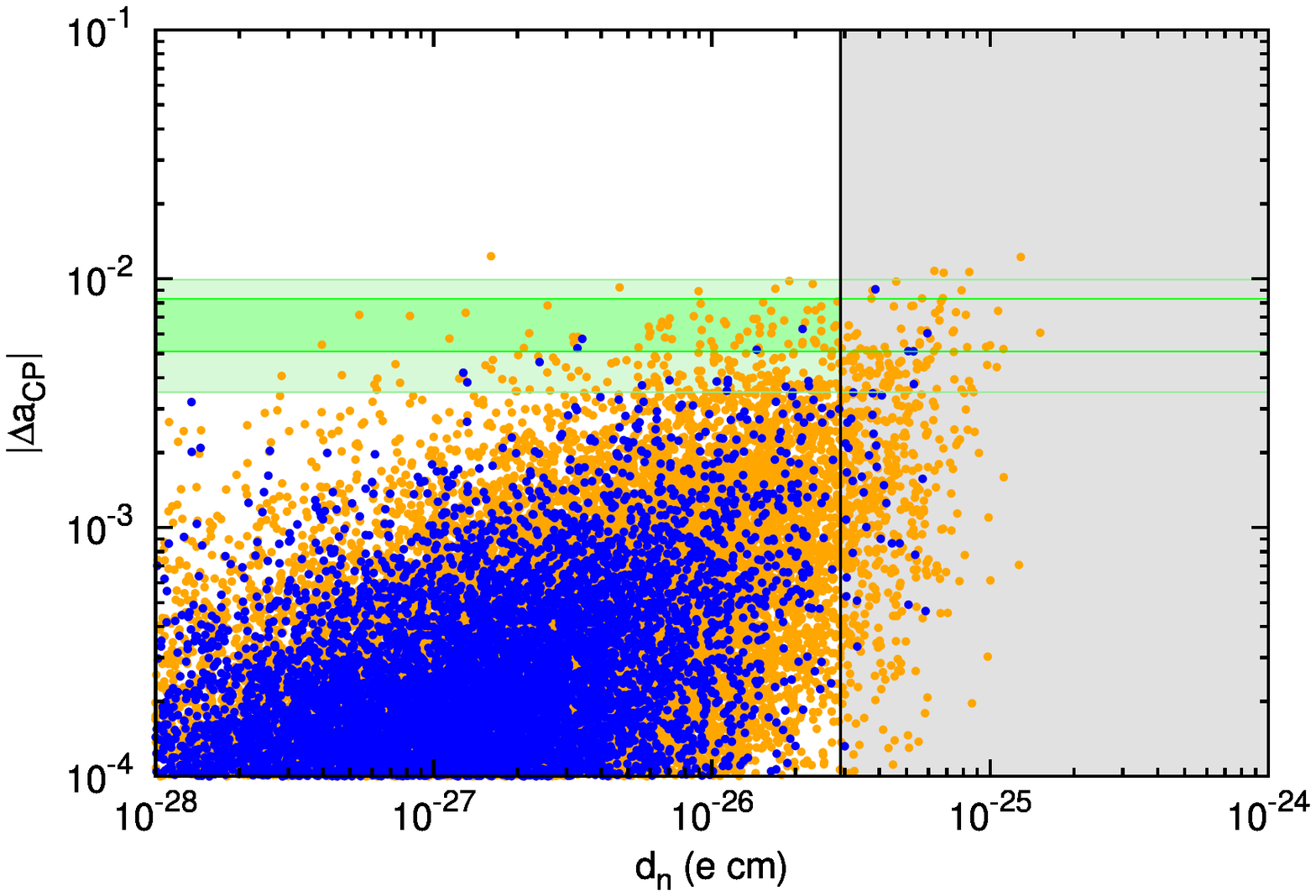}
\caption{\label{fig:u1} $|\Delta a_{CP}|$ versus $|a_{(\rho,\omega)\gamma}|$ (left) and $d_n$ (right) in the FGM$_U$ + $U(1)$ model
for a wide scan of the parameters (see the text for details). The blue points correspond to $m_h>123$ GeV. The green (dark green) 
band represents the 2$\sigma$ (1$\sigma$) experimental range. The grey shaded area is excluded by neutron EDM searches.}
\end{figure}

To illustrate what discussed above, we show a numerical computation of $\Delta a_{CP}$
in the FGM$_U$ + $U(1)$ model. The gauge mediation parameters are varied as in the scan presented 
in Section 4, see Eq.~(\ref{eq:flavor-scan}). The structure of $\lambda_U$ is dictated by the
$U(1)$ symmetry as in Eq.~(\ref{eq:u1}), taking random $\ord{1}$ coefficients (between 0.3 and 3) 
and phases. A possible hadronic enhancement of $\Delta a_{CP}$ (up to a factor of 3) is taken into account.
In Fig.~\ref{fig:u1} we plot the resulting $|\Delta a_{CP}|$ vs. $|a_{(\rho,\omega)\gamma}|$ (left) 
and $d_n$ (right). As before the blue points correspond to $m_h>123$ GeV. 
We see that in the model a potentially large SUSY contribution to $|\Delta a_{CP}|$ 
is not excluded by the bound from the neutron EDM, as discussed above.

Finally we briefly comment on the leptonic sector. Including the new coupling $\Delta W = \lambda_E L E \overline{\Phi}_{H_d}$ with $\lambda_E \sim y_E$, by a spurion analysis we find that the flavor structure of the soft terms has the form
\begin{equation}
A_E \sim  \lee \led \ye + \ye \led \lee,
\end{equation}
\begin{equation}
\Delta \tilde{m}^2_L  \sim  \lee \led,~\qquad  \Delta \tilde{m}^2_E  \sim  \led \lee,
\end{equation}
If we restrict to the case where a single $U(1)$ symmetry controls 
the size of the flavor couplings, and take for the unitary rotations in the charged lepton sector the rough estimate $(V^E_{L,R})_{ij} \approx \sqrt{m^E_i/m^E_j}$ for $i \le j$, we find
\begin{equation}
(\delta^e_{LL})_{ij} \sim (\delta^e_{RR})_{ij} \sim  y_{\tau}^2 \frac{\sqrt{m^E_i m^E_j}}{m_\tau},
~\qquad
(\delta^e_{LR})_{ij} \sim  \frac{(m^E_i + m^E_j ) A}{{\tilde m}_L{\tilde m}_E} y_{\tau}^2
\frac{\sqrt{m^E_i m^E_j}}{m_\tau}.
\end{equation}
Again for light generations the effective MIs generated from triple or double products
can be dominant, with the leading contribution given by
\begin{eqnarray}
(\delta^e_{LR})_{ij}^{eff} &\sim&
\frac{ m_\tau \left( A - \mu^* \tan\beta \right) }{ {\tilde m}_L{\tilde m}_E }
y_{\tau}^4 \frac{\sqrt{m^E_i m^E_j}}{m_\tau} ~\qquad i,j = 1,2.
\end{eqnarray}
The most stringent constraints that arise from $\mu\to e\gamma$ and the electron
EDM~\cite{Masina:2002mv,Paradisi:2005fk} can be naturally satisfied for moderate
$\tan\beta$ values even for ${\tilde m}_L \sim {\tilde m}_E \sim 200~$GeV.
Therefore, the FGM scenario represents a concrete example for those classes of models
where the muon $g-2$ anomaly can be naturally accounted for while keeping under control other
dipole transitions such as $\mu\to e\gamma$ and the electron EDM~\cite{Giudice:2012ms}.

\section{Conclusions}

Among the many candidates for a SUSY breaking mechanism, Gauge Mediation provides an elegant and
very predictive framework. This scenario naturally realizes the Minimal Flavor Violation paradigm,
and therefore seems to be favored by the current flavor data where no convincing non-SM signals
have been observed so far.
On the other hand, minimal realizations of GMSB are now seriously challenged by the Higgs boson
discovery at the LHC, since  they can account for $m_h\approx 126$ GeV only at the price of large
fine-tuning and a spectrum beyond the reach of the LHC.

This has recently motivated several attempts to extend the minimal GMSB framework by direct couplings
of GMSB messengers to MSSM fields~\cite{Shadmi1, Y1, Y2, Kang, Craig, Babu, Shadmi2, Ray, EvansShih}.
While most of this models preserve the MFV structure of Minimal Gauge Mediation, we found particularly
interesting the setup considered in Ref.~\cite{Shadmi1}, dubbed ``Flavored Gauge Mediation" (FGM).
In this work the authors have proposed couplings of the messengers to the MSSM matter fields, which
resemble the MSSM Yukawa couplings and are therefore assumed to be controlled by the same underlying
mechanism that explains flavor hierarchies.
This framework can be easily motivated, provided that the underlying theory of flavor treats
GMSB messengers and MSSM Higgs fields in the same way, for example as a consequence of a
flavor symmetry under which messengers and Higgs have the same quantum numbers.

While in Ref.~\cite{Shadmi1, Shadmi2} the authors concentrated mainly on the  implications for
the SUSY spectrum and the slepton flavor sector, in this work we have studied the general flavor structure of this framework in great detail.
We have shown that this scenario gives rise to an interesting pattern of flavor violation
that goes in a controlled way beyond MFV, in which the dominant effects enter through
A-terms, i.e. LR mass insertions, while effects from LL and RR mass insertions are very 
efficiently suppressed (see Table~\ref{table}, \ref{table2}).
This strong suppression is reminiscent of what happens in the case of wave function 
renormalization~\cite{Pokorski, Isidori} or Partial Compositeness~\cite{Nomura, Lodone}, despite the
underlying flavor model can be a simple U(1) flavor model (which in the context of Gravity Mediation
typically suffers from strong $\Delta S=2$ constraints~\cite{Nir, Cetal}). In FGM this extra
suppression originates in the loop origin of soft terms, which acts
precisely as a wavefunction suppression. Indeed LL and RR mass insertions, as well as LR mass insertions in the up sector, are roughly suppressed as in Partial Compositeness. Interestingly,
the LR mass insertions in the down sector are additionally suppressed by $y_b^2$ (which becomes 
strong in the low $\tan\beta$ regime) and therefore dipole transitions such as hadronic EDMs and $\eps^{\prime}/\eps$ are better controlled than in Partial Compositeness. Still, an important difference is that in FGM the spectrum is completely calculable in terms of few parameters.
Another virtue of this model, as opposed to PC and $U(1)$ flavor models in the context of Gravity Mediation, is the additional built-in protection against flavor-blind CPV phases stemming from the
GMSB loop structure.

This setup therefore allows the implementation of SUSY flavor models (in particular $U(1)$ models) 
with a built-in suppression of $\Delta F=2$ effects and flavor-violating effects mainly arising from
the LR MIs in the up-sector. This naturally realizes the ``disoriented" A-term scenario~\cite{GIP}
and thus provides an ideal framework to account for the observed direct CP violation in charm decays.

Concerning the phenomenology of this model, we summarize here our main findings:
\begin{itemize}
\item With respect to minimal gauge mediation, the spectrum of the model is basically controlled
by a single new parameter of the size of the top Yukawa. For a broad range of this parameter
the SUSY spectrum is strongly modified with respect to minimal GMSB, with either light stops
or light first generation squarks and gluinos that are potentially observable at the LHC
(see Fig.~\ref{specA}, \ref{spec122}).
\item The lightest Higgs boson mass $m_h\approx 126$ GeV can be accounted for by stop masses
around 1 TeV (see Fig.~\ref{spec122}). On the other hand, the Higgs boson properties remain basically
SM-like. In particular, in spite of the relatively light stau (with large left-right mixing)
and heavy higgsinos, we find only a few per-cent enhancement in $h\to\gamma\gamma$
since $m_{{\tilde\tau}_1}\gtrsim 200~$GeV (see Fig.~\ref{spec122}).
\item The SUSY contribution to $(g-2)_\mu$ can be easily as large as $(1\div 2)\times 10^{-9}$
(see Fig.~\ref{gm2}), thus reducing significantly the $\sim 3.5 \sigma$ discrepancy between the
SM prediction and the experimental value $\Delta a_\mu =a_\mu^{\mysmall \rm EXP}-a_\mu^{\mysmall
\rm SM} = 2.90 (90) \times 10^{-9}$~\cite{bnl, Jegerlehner:2009ry, HLMNT11, DHMZ11}.
\item Since flavor-violating effects mainly arise in the LR up-sector, we can easily explain
the observed direct charm-CPV $\Delta a_{CP} = -(0.68 \pm 0.15)\%$,
while being compatible with all $\Delta F=2$ and EDM bounds (see Fig~\ref{fig:DaCP}, \ref{fig:u1}).
Yet, it is easy to generate sizable NP effects for $\epsilon_K$ (but not for $B_{d,s}$ mixing) 
which can improve the UT fit~\cite{UTfit}.
\item In the lepton sector, we find that LFV processes like $\mu\to e\gamma$ and the electron 
EDM are naturally under control with typical predictions lying within the expected experimental 
resolutions.
\end{itemize}
In conclusion, we have discussed a concrete model within a GMSB framework departing
in a controlled way from the MFV paradigm, along the lines of Ref.~\cite{Shadmi1}. 
While this model is able to satisfy naturally all the current bounds from direct and 
indirect searches, it can also accommodate the few possible hints of New Physics.

\section*{Acknowledgements}
We would like to thank Gian F. Giudice and Gino Isidori for useful discussions.
We also thank the CERN Theory Group and the University of Warsaw for kind hospitality,
where this work was partially completed. This research was done in the context of 
the ERC Advanced Grant project ''FLAVOUR''(267104) and was partially supported (R.Z) 
by the TUM Institute for Advanced Study.

%%%%%%%%%%%%%%%%%%%%%%%%%%%%%%%%%%%%%%%%%%%%%%%%%%%%%%

\end{document}